\documentclass[journal=jpcafh,manuscript=article]{achemso}

\usepackage{chemformula} 
\usepackage[T1]{fontenc} 

\author{Jeffrey D. Mills}
\email{jeffrey.mills.4@spaceforce.mil; jdmills_chemphys@hotmail.com}
\affiliation[AFRL]
{Air Force Research Lab., 10 E. Saturn Blvd., Edwards AFB, CA
  93524, United States}

\title[A]
{Applications of \\
  the Spectral Theory of Chemical Bonding \\
  to Simple Hydrocarbons
}

\begin{document}







\begin{abstract}

  The finite-basis, pair formulation of the Spectral Theory of
  chemical bonding is briefly surveyed.  Solutions of the
  Born-Oppenheimer polyatomic Hamiltonian totally antisymmetric in
  electron exchange are obtained from diagonalization of an aggregate
  matrix built up from conventional diatomic solutions to
  atom-localized problems.  A succession of transformations of the
  bases of the underlying matrices and the unique character of
  symmetric orthogonalization in producing the archived matrices
  calculated ``once-of-all'' in the pairwise-antisymmetrized basis are
  described.  Application is made to molecules containing hydrogens
  and a single carbon atom.  Results in conventional orbital bases are
  given and compared to experimental and high-level theoretical
  results.  Chemical valence is shown to be respected and subtle
  angular effects in polyatomic contexts are reproduced.  Means of
  reducing the size of the atomic-state basis and improve the fidelity
  of the diatomic descriptions for fixed basis size, so as to enable
  application to larger polyatomic molecules, is outlined along with
  future initiatives and prospects.
  
\end{abstract}

\section{Introduction}

That molecules are composed of interacting atoms was an important
conceptual touchstone in the development of classical chemistry.  From
this perspective it is perhaps surprising that most contemporary
quantum-chemical methods do not make direct use of this insight but
construct molecules from electrons and nucleii, instead.  In
particular the initial step often involves forming, from a number of
localized basis functions, one-electron orbitals delocalized across
the entire molecule under a mean-field-type, electron-electron
interaction.  Subsequent correction of this solution can then be
regarded as introducing partial relocalization.  {\it Post hoc}
interpretations of these solutions, often involving a
localization-approximating unitary transformation or other
mathematical manipulations, can then be used in at attempt to gain
insights in terms of atoms and their interactions%
\cite{ballp11a,
  bader90ax,blanm05a,frane06a,rychj86a,parrr05a,ruedk13a,ruedk22a,herbj19a}.
(No effort is made here to completely review this literature, but only
to provide a starting point for interested readers.)

This publication describes results of the application of one variety
of a class of methods referred to as the Spectral Theory of chemical
bonding,%
\cite{langp96a,langp02a,langp02b,langp04a,langp04b,langp08a,
bennm09a,millj16a,langp18a,millj18a}
which seeks to directly construct quantum-mechanical descriptions of
polyatomic molecules from the characteristics of the constituent
atoms, namely their electronic eigenspectrum and the state spectra of
interacting pairs.  These methods differ from some others with similar
motivations (also not an exhaustive citation list)%
\cite{moffw51b,moffw54a,ellif63a,ellif63b,ellif64a} in specific
implementation details.

Principal advantages of the Spectral Theory will be shown to include
relief from the need to calculate two-electron integrals over three or
four centers and the ``once for all'' determination of reference
diatomic matrices ready for universal application to the pairs in any
polyatomic molecule which contains the two atoms.  Much effort in this
(and subsequent publications) is geared toward overcoming the severe
size-scaling which will be shown to be exponentially dependent on the
number of atoms in the molecule and polynomially dependent upon the
size of the state space used to describe the atoms.  The overall
motivation for the Spectral Theory of chemical bonding is the
possibility of recovering, for an N-atom molecule, a significant
portion of the N-excitation CI energy, all from archived data only
single-excitation in the atoms and (atom-centered) double-excitation
in the diatoms.

It is the purpose of this report to itemize the steps of the
pairwise-antisymmetrized form of the Spectral Theory in constructing
polyatomic wave functions from atomic states via the mediation of
totally antisymmetric treatments of the unique interacting pairs of
atoms in the molecule.  After a description of the theoretical
procedures, as currently deployed, there is an account of some results
of the application of the theory to simple paradigmatic hydrocarbons
containing a single carbon atom and two means by which the atomic
state space might be kept relatively small while retaining accuracy.

\section{Theoretical methods}

\subsection{Summary of theoretical/computational development}

\begin{figure}
  \includegraphics[width=0.7\textwidth,angle=-90.]{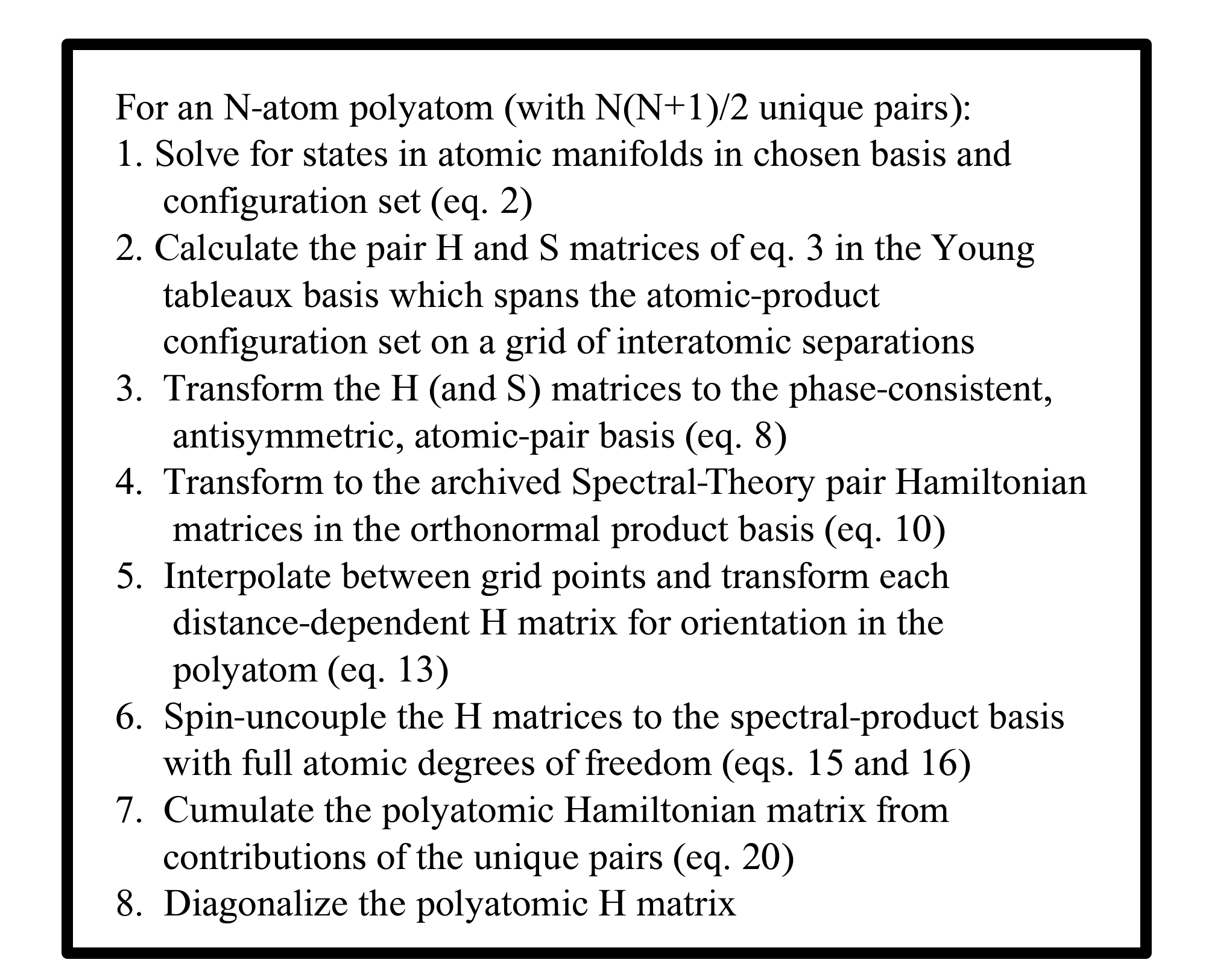}
  \caption{Outline of the steps of the finite-basis, pairwise-antisymmetrized form
    of the Spectral Theory}
  \label{fgr:STout}
\end{figure}

Much of this discussion follows the progress and notation of
ref.~\citenum{millj16a}, especially sections 2 and 3, but focuses on
the current flow and set of options to distinguish them from the
greater number of possibilities described in previous publications.
In addition a few new approaches or deviations from what are described
there are explained in more detail. The overall organization is
motivated by a ``building up'' from the more conventional
quantum-chemical components toward the more unique aspects of the
Spectral Theory.  A brief outline of the steps is given in
Figure~\ref{fgr:STout} for reference.

The overall goal of this section is to justify the construction of
solutions of the polyatomic Born-Oppenheimer, quantum-chemical
Hamiltonian which are totally antisymmetric in electron exchange and
built from finite-basis treatments of all the molecular pairs.  These
diatomic basis components are themselves totally antisymmetric
products of atomic states.  Previous publications have referred to
this as the ``pairwise-antisymmetrized, finite-subspace
representation.''

The Spectral Theory development starts from a type of valence-bond
theory especially amenable to localized, in this case atomic-centered,
calculations.  Sometimes referred to as ``nonorthogonal valence-bond
theory'' (NOVB) or ``multi-configurational valence bond theory'' (MCVB) this
configuration-interaction-based approach is implemented in the CRUNCH
suite of computational chemistry codes.\cite{gallg02ax} Putting aside
implementation details, it can be considered as equivalent to the more
common molecular-orbital, configuration-interaction (MOCI) approach
but with the additional flexibility that the one-electron orbitals
which comprise the configurations need not be orthogonal, but can be
strictly localized and even determined by the solution of atomic
problems.

\subsection{Valence-bond methods for atomic eigensolutions}

In the CRUNCH codes configurations for the atomic CI calculations are
space-symmetry-adapted constellations of Young standard
tableaux\cite{gallg02ax,chisc76ax,wilss03c} composed of the
one-electron orbitals and defined by
the NPN spin representation of the symmetric group. These formally
spin-free functions in the spatial coordinates $\boldsymbol{i}$ of
atom $\alpha$ are labeled
$\boldsymbol{\Phi}_{\bf yt}^{(\alpha)} (\boldsymbol{i})$.  (In
previous Spectral-Theory publications finite-basis matrices and basis
functions were notated with a ``tilde,'' but since only finite-bases
are considered here that convention is abandoned.)  Because the
angular portion of the one-electron functions are complex spherical
harmonics, the multi-electron configuration functions each have good
spin (S), angular-momentum-projection ($\mathrm{M_l}$), and parity (P
for g/u) quantum numbers and the configurations can be distinguished
by their different orbital occupations.  Orbital occupations are
chosen such that complete manifolds of $\mathrm{M_l}$ functions for
each desired atomic L value are spanned.

Strictly, of course, the computations don't use or express these
functions directly, but deliver matrix elements between functions with
different orbital occupations and solve matrix equations involving the
elements.  Because the standard tableau functions (as employed in the
current code suite) are not orthonormal, wave functions must be
obtained by solution of the generalized-eigenvalue problem.


\begin{equation}                                                                
  {\bf H}_{\bf yt}^{(\alpha)} \cdot {\bf V}_{{\bf H}_{\bf yt}}^{(\alpha)} =     
  {\bf S}_{\bf yt}^{(\alpha)} \cdot {\bf V}_{{\bf H}_{\bf yt}}^{(\alpha)}\cdot  
  {\bf E}^{(\alpha)}                                                            
  \label{eqn:atgenE}                                                            
\end{equation}

\noindent ${\bf H}_{\bf yt}$ and ${\bf S}_{\bf yt}$ are the
Hamiltonian and overlap matrices in the Young Tableau basis ($\bf yt$)
and ${\bf E}$ is a diagonal matrix of the eigenvalues.  Orthonormal
solutions are then obtained from the Young-tableau basis according to:


\begin{equation}                                                                
  \boldsymbol{\Psi}^{(\alpha)}(\boldsymbol{i}) =                                
  \boldsymbol{\Phi}_{\bf yt}^{(\alpha)}(\boldsymbol{i}) \cdot                   
{\bf V}_{{\bf H}_{\bf yt}}^{(\alpha)}\label{eqn:atsol}                          
\end{equation}

\noindent In addition to the quantum numbers of the configurational
basis in which the solution is sought (S, $\mathrm{M_l}$, and P), the
eigensolutions, because of interactions in the atomic Hamiltonian,
manifest additional good quantum numbers.  Total orbital angular
momentum, L, and an index e which is the order of the energy among
solutions with the same values of the other quantum numbers (made
unique, as necessary, in the case of accidental degeneracy) are
manifested.

\subsection{Diatomic solutions in a Young tableau basis}

The CRUNCH codes are also used to solve the diatomic problems.  The
atoms ($\alpha$ and $\beta$) are separated along their quantization axis so
the one-electron spatial orbitals can be identical to those used for
the atoms but placed separately on each nucleus.  Orbitals on
different atoms are therefore not orthogonal and neither are
constellations of the diatomic Young tableaux which result.  The
configurations have good quantum numbers associated
with the total diatomic spin ($\mathrm{S}_{\alpha\beta}$) and the
projection of the total orbital angular momentum along the axis of
nuclear separation ($\lambda_{\alpha\beta}^\prime$) and can be
distinguished or labeled by their orbital occupations.  (The
``prime'' indicates that the quantization axis for angular momentum is
different from the global polyatomic quantization axis which will be
adopted later.)  The CRUNCH codes are capable of forming
symmetry-adapted constellations appropriate for homonuclear diatomic
molecules in the $D_{\infty h}$ point group.  However, because of the
transformations described in the next section, the present
calculations employ multi-electron configurations only constrained to
$C_{\infty v}$.  In addition the +/- functions
(distinguished by their eigenvalues with respect to reflection through
planes containing the internuclear axis) are lumped together as a
matter of practice.  The sets of orbital occupations are chosen to
encompass all combined possibilities of the atomic calculations
described in the previous section and so the diatomic treatments are
completely size-consistent.

Although not of primary interest to the current investigation, for a
diatom ($\alpha,\beta$) with electrons ($\boldsymbol{i}$,
$\boldsymbol{j}$), eigenvalues
${\bf E}^{(\alpha,\beta)}(R_{\alpha\beta})$ may be obtained by solving
the generalized-eigenvalue problem:


\begin{equation}
  {\bf H}_{\bf yt}^{(\alpha,\beta)}(R_{\alpha\beta}) \cdot
  {\bf V}_{{\bf H}_{\bf yt}}^{(\alpha,\beta)}(R_{\alpha\beta}) =
  {\bf S}_{\bf yt}^{(\alpha,\beta)}(R_{\alpha\beta}) \cdot
  {\bf V}_{{\bf H}_{\bf yt}}^{(\alpha,\beta)}(R_{\alpha\beta})\cdot
  {\bf E}^{(\alpha,\beta)}(R_{\alpha\beta})
  \label{eqn:diytsolv}
\end{equation}

\noindent Orthonormal
eigenfunctions in the constellation basis are then constructed as:


\begin{equation}
\boldsymbol{\Psi}^{(\alpha,\beta)}(\boldsymbol{i},\boldsymbol{j};
R_{\alpha\beta}) = 
\boldsymbol{\Phi}_{\bf yt}^{(\alpha,\beta)}(\boldsymbol{i},\boldsymbol{j})\cdot
{\bf V}_{{\bf H}_{\bf yt}}^{(\alpha,\beta)}(R_{\alpha\beta})
\label{eqn:diytsolu}
\end{equation}

\noindent The interactions in the diatomic Hamiltonian add parity, P (for
g/u (homonuclear)), reflection symmetry, $\sigma$ (for +/-), and an
energy index, e, to the good quantum numbers of the CI basis functions
already listed.

More directly relevant as the starting point in moving toward the
polyatomic Spectral Theory are the Hamiltonian and overlap
matrices of eq.~\ref{eqn:diytsolv} in the configuration basis.  The
linear transformations of this basis
which are formulated as transformations of the matrices
represented computationally, make up the bulk of the remainder of this
section.

Before proceeding, it is perhaps appropriate to discuss in more detail
the implications of and to distinguish between what are referred to as
``good quantum numbers'' and ``basis-set labels.''  In the context of
building up and transforming matrices for the Spectral Theory, and a
direct implication of the common usage of this phrase in quantum
mechanics, good quantum numbers define Hamiltonian and overlap
matrix sub-blocks which, if the underlying subbases are concatenated,
have no non-zero matrix elements between them.  With proper basis-set order, the supermatrices are block
diagonal.  Basis set labels are used to definitively identify
basis functions, especially with respect to the character of
the atomic components of the diatomic functions.  Within a matrix block
involving states with a common set of quantum numbers there are
generally interconnecting non-zero matrix elements between functions with
different labels.

\subsection{Phase-consistent, atomic-pair basis}

The chain of linear transformations of the Spectral Theory begins with
modifying the solutions of the atomic problem.  The Young-tableaux
basis for each atom is transformed to the atomic eigenstate basis of
eq.~\ref{eqn:atsol}.  In addition, because of the need to build up
polyatomic functions with a phase consistent among the
fragments, the arbitrary (with respect to angular-momentum algebra)
phase (sign) introduced into ${\bf V}_{{\bf H}_{\bf yt}}^{(\alpha)}$
by the numerical diagonalization of eq.~\ref{eqn:atgenE} is
remedied.

Each of the 2L+1 eigenstates belonging to an L manifold are separately
considered.  Those with the 2L largest values of $\mathrm{M_l}$ are
replaced sequentially as follows.  Using eq.~\ref{eqn:atsol} the
eigenfunction with $\mathrm{M_l}$ = -L can be written as an expansion
of the CI basis functions.  Each of its components can
then be analytically subjected to orbital- and state- raising
operators\cite{edmoa74ax} with tableau-rearrangement algorithms
allowing re-expression in terms of a linear combination of standard
tableaux of unit-increased $\mathrm{M_l}$.  This new function then
replaces the phase-arbitrary initial solution and the process is
repeated to phase-rectify the remaining coefficients in each L manifold.
Because of the known phase relationship between the one-electron
orbitals (preserved in the orbital raising operators), this new
manifold of eigensolutions possesses within itself a sign consistency
which will ultimately allow the diatomic pair functions to be
subjected to a coordinate rotation (eq.~\ref{eqn:rotst}) which will
mix states of differing atomic $\mathrm{M_l}$ but the same L (and
other quantum numbers).  This is possible in spite of the fact that
the relative phase between different L manifolds (because of the
arbitrary phase between their root, diagonalizer-generated
$\mathrm{M_l}$ = -L states) remains arbitrary and unknown.

The phase-consistent ($\bf pc$) atomic functions which result can be denoted:



\begin{equation}
  \boldsymbol{\Psi}_{\bf pc}^{(\alpha)}(\boldsymbol{i}) = 
  \boldsymbol{\Phi}_{\bf yt}^{(\alpha)}(\boldsymbol{i}) \cdot
{\bf V}_{{\bf H}_{\bf pc}}^{(\alpha)}\label{eqn:atpc}
\end{equation}

\noindent In addition to being phase consistent, this set of atomic
eigenfunctions retain the good quantum numbers L, $\mathrm{M_l}$, S,
P, and N of the functions of eq.~\ref{eqn:atsol}.

The second step involves incorporation of what are referred to as the
subduction coefficients\cite{lomoj59ax,clavp71a,mcwer89ax}. These
allow each diatomic Young tableau, which is totally antisymmetric with
respect to electron exchange, to be expressed exactly as a likewise
totally antisymmetric linear combination of direct products of atomic
Young tableaux.  Thus the diatomic basis of
eq.~\ref{eqn:diytsolu} can be transformed to an atomic-product basis
spanning the same space as:


\begin{equation}
  \boldsymbol{\Phi}^{(\alpha\otimes\beta)}(\boldsymbol{i},\boldsymbol{j}) = 
  \boldsymbol{\Phi}_{\bf
    yt}^{(\alpha,\beta)}(\boldsymbol{i},\boldsymbol{j})\cdot 
  {\bf D}_{\bf sub}^{(\alpha,\beta)}\label{eqn:subd}
\end{equation}

\noindent The real matrix ${\bf D}_{\bf sub}^{(\alpha,\beta)}$,
independent of atomic separation, includes non-zero coefficients for
all spin-couplings between $\mathrm{S}_{\alpha}$ and
$\mathrm{S}_{\beta}$ necessary to yield $\mathrm{S}_{\alpha\beta}$ in
a common set of orbitals.

Finally combining the coefficients of the phase-consistent atomic
eigenstates ${\bf V}_{{\bf H}_{\bf pc}}^{(\alpha)}$ and
${\bf V}_{{\bf H}_{\bf pc}}^{(\beta)}$ with the subduction
coefficients ${\bf D}_{\bf sub}^{(\alpha,\beta)}$ into a single
orthogonal matrix ${\bf M}_{\bf pc}^{(\alpha,\beta)}$, one obtains the
atomic-state pair basis which is referred to in
ref.~\citenum{millj16a} as the valence-bond basis:


\begin{equation}
  \boldsymbol{\Phi}_{\bf
    vb}^{(\alpha,\beta)}(\boldsymbol{i},\boldsymbol{j}) \equiv 
  \{\boldsymbol{\Phi}_{\bf
    pc}^{(\alpha)}(\boldsymbol{i})\otimes
  \boldsymbol{\Phi}_{\bf
    pc}^{(\beta)}(\boldsymbol{j})\} =
  \boldsymbol{\Phi}_{\bf
    yt}^{(\alpha,\beta)}(\boldsymbol{i},\boldsymbol{j})\cdot
  {\bf M}_{\bf pc}^{(\alpha,\beta)}
  \label{eqn:vbbas}
\end{equation}

\noindent Totally antisymmetric combinations of the phase-consistent atomic
eigenstates can thus provide a basis for solution of the diatomic CI
problem.  The functions in this basis each have good diatomic quantum
numbers $\mathrm{S}_{\alpha\beta}$ and $\lambda_{\alpha\beta}^\prime$
which are represented collectively by $(\alpha,\beta)$ in the left
most symbol.  Also implicit in this notation are the basis labels for
each of the atomic states of which it is composed
($\mathrm{L}_\alpha$, $\mathrm{M}_{\mathrm{l}\alpha}^\prime$,
$\mathrm{S}_\alpha$, $\mathrm{P}_{\alpha}$, $\mathrm{e}_\alpha$;
$\mathrm{L}_\beta$, $\mathrm{M}_{\mathrm{l}\beta}^\prime$,
$\mathrm{S}_\beta$, $\mathrm{P}_\beta$, $\mathrm{e}_\beta$).  In
following sections this diatomic basis will be subjected to linear
transformations which represent a change in quantum numbers and/or
atomic state labels.  The more generic representation $(\alpha,\beta)$
will sometimes be replaced by others which emphasize the change in
a particular characteristic of the basis.  To do so strictly, for every degree of
freedom, would excessively proliferate new symbols/superscripts.
Nevertheless the good quantum numbers and distinguishing labels of
transformed bases will be emphasized in the text accompanying the
equations.

Numerically it is the diatomic Hamiltonian and overlap
matrices rather than the basis functions, {\it per se}, which are
subjected to this transformation as, for example:


\begin{equation}
  {\bf H}_{\bf vb}^{(\alpha,\beta)}(R_{\alpha\beta}) =
  {\bf M}_{\bf pc}^{(\alpha,\beta)\,t}\cdot
{\bf H}_{\bf yt}^{(\alpha,\beta)}(R_{\alpha\beta})\cdot
{\bf M}_{\bf pc}^{(\alpha,\beta)}
  \label{eqn:yttovb}
\end{equation}

\noindent For simplicity the diatomic basis has not been notated as
being dependent upon internuclear separation.  It is in the
calculation of matrix elements from the one- and two-electron
integrals over functions located on the atoms that the R dependence
arises.  For convenience, the overlap and Hamiltonian
matrices for each separation are now subjected to normalization of the
underlying basis.

It is worth noting that as the separation of the atoms approaches
infinity, the underlying diatomic basis becomes orthonormal; the
overlap matrix becomes the identity and the Hamiltonian matrix becomes
diagonal with sums of atomic energies as the only non-zero elements.
For finite separations both matrices are generally dense.

It should further be emphasized that this new basis spans the same
space as the diatomic Young tableau basis and so it yields precisely
the same eigenvalues when these matrices are subjected to
diagonalization.  The eigenvectors would now be given in the basis of
antisymmetric products of atomic eigenstates ({\it c.f.}
eq.~\ref{eqn:diytsolu}).

\subsection{Orthogonalized diatomic matrix archive}

As is typical in quantum chemistry codes, the CRUNCH suite focuses on
calculating the active electronic energy.  At this point in the
Spectral Theory development (although for simplicity it will not be
indicated by a change of the notation) the nuclear repulsion and any
frozen-core electronic energy are added back into the Hamiltonian
matrices so that they describe total diatomic energies relative to
separated electrons and nucleii.  In the context of the non-orthogonal
basis, this involves replacing
${\bf H}_{\bf vb}^{(\alpha,\beta)}(R_{\alpha\beta})$ with:


\begin{equation}
{\bf H}_{\bf vb}^{(\alpha,\beta)}(R_{\alpha\beta})+(E_{\mathrm{nuc}}+E_{\mathrm{core}})
{\bf S}_{\bf vb}^{(\alpha,\beta)}(R_{\alpha\beta})
  \label{eqn:addcor}
\end{equation}

Now for each distance and set of diatomic quantum numbers the overlap
matrix is diagonalized and the Hamiltonian matrix is transformed to
reflect an orthogonalization of the underlying basis.


\begin{equation}
{\bf H}_{\bf d}^{(\alpha,\beta)}(R_{\alpha\beta}) =
{\bf A}_{\bf so}(R_{\alpha\beta})^{t} \cdot
{\bf H}_{\bf vb}^{(\alpha,\beta)}(R_{\alpha\beta}) \cdot
{\bf A}_{\bf so}(R_{\alpha\beta})
  \label{eqn:sotrans}
\end{equation}

\noindent Symmetric orthogonalization\cite{lowdp70a} is used and so
the transformation matrix is given by:


\begin{equation}
{\bf A}_{\bf so}(R_{\alpha\beta}) =
{\bf U}_{\bf S_{vb}}^{(\alpha,\beta)}(R_{\alpha\beta}) \cdot
{\bf s}_{\bf vb}^{(\alpha,\beta)}(R_{\alpha\beta})^{-1/2} \cdot
{\bf U}_{\bf S_{vb}}^{(\alpha,\beta)}(R_{\alpha\beta})^t
  \label{eqn:somat}
\end{equation}

\noindent ${\bf U}_{\bf S_{vb}}^{(\alpha,\beta)}(R_{\alpha\beta})$ is the matrix
which diagonalizes
${\bf S}_{\bf vb}^{(\alpha,\beta)}(R_{\alpha\beta})$ and
${\bf s}_{\bf vb}^{(\alpha,\beta)}(R_{\alpha\beta})^{-1/2}$ is a
diagonal matrix with the inverse square roots of the corresponding
eigenvalues on the diagonal.  The transformation matrix
${\bf A}_{\bf so}(R_{\alpha\beta})$ is generally dense and so diatomic
functions with different atomic basis labels are extensively mixed,
especially at smaller separation. Furthermore, it should be noted that
in symmetric orthogonalization arbitrary signs in
${\bf U}_{\bf S_{vb}}^{(\alpha,\beta)}(R_{\alpha\beta})$ which arise
from diagonalization are neutralized as they appear in pairs in
eq.~\ref{eqn:somat} and so
the phase-consistent nature of the basis is preserved.  If the
functions in the underlying nonorthogonal basis are linearly
independent (to within numerical precision), the same energies
(accounting for the differences in nuclear and core-electron
energies) as the original Young-Tableau basis (eq.~\ref{eqn:diytsolv})
are obtained by solution of the regular eigenvalue problem:


\begin{equation}
  {\bf H}_{\bf d}^{(\alpha,\beta)}(R_{\alpha\beta}) \cdot
  {\bf V}_{{\bf H}_{\bf d}}^{(\alpha,\beta)}(R_{\alpha\beta}) =
  {\bf V}_{{\bf H}_{\bf d}}^{(\alpha,\beta)}(R_{\alpha\beta})\cdot
  {\bf E}^{(\alpha,\beta)}(R_{\alpha\beta})
  \label{eqn:distsol}
\end{equation}

Only the transformed Hamiltonian matrices
${\bf H}_{\bf d}^{(\alpha,\beta)}(R_{\alpha\beta})$ in the
antisymmetrized, orthonormal product basis are carried forward
hence. In practice matrices generated on a fixed grid of
$R_{\alpha\beta}$ values are archived along with cubic spline
coefficients to interpolate corresponding matrix elements at distances
between the grid points.  Together these comprise the diatomic
information which need only be calculated ``once for all'' with later
application to different polyatomic systems or altered polyatomic
geometries.

\subsection{Orientation of pair in polyatom}

Each of the pairs in a polyatomic molecule with vector separation
$\boldsymbol{R}_{\alpha\beta}$ has a magnitude of internuclear
separation $R_{\alpha\beta}$ and a lab-frame orientation
$\boldsymbol{\hat{R}}_{\alpha\beta}$.  After interpolating to a
separation between grid points, if necessary, the Hamiltonian
matrix for each pair is transformed to reflect the orientation of the
diatom in the polyatom.  Specifically the quantization direction for
the projection of orbital angular momentum is rotated to a global
orientation common to all pairs in the polyatom.


\begin{equation}
  {\bf H}_{\bf d}^{(\alpha,\beta)}(\boldsymbol{R}_{\alpha\beta}) =
  {\bf D}^{(\alpha,\beta)}(\boldsymbol{\hat{R}}_{\alpha\beta})^\dagger\cdot
{\bf H}_{\bf d}^{(\alpha,\beta)}(R_{\alpha\beta})\cdot
{\bf D}^{(\alpha,\beta)}(\boldsymbol{\hat{R}}_{\alpha\beta})
  \label{eqn:rotst}
\end{equation}

\noindent In terms of quantum numbers and labels, this new basis loses
its orbital angular momentum projection quantum number in the new
frame but retains the diatomic spin quantum number
$\mathrm{S}_{\alpha\beta}$.  In addition each basis function is
labeled by the atomic eigenfunctions which compose it,
$\mathrm{L}_\alpha$, $\mathrm{M}_{\mathrm{l}\alpha}$,
$\mathrm{S}_\alpha$, $\mathrm{P}_{\alpha}$, $\mathrm{e}_\alpha$;
$\mathrm{L}_\beta$, $\mathrm{M}_{\mathrm{l}\beta}$,
$\mathrm{S}_\beta$, $\mathrm{P}_\beta$, $\mathrm{e}_\beta$, where the
removal of the ``prime'' indicates the rotation to a common polyatomic
axis.  Aside from the change from scalar to vector internuclear
spacing, this will not be further reflected in the symbol for the
matrix itself.

The diatomic transformation matrix is composed from direct products of
the Wigner rotation matrices for the two atoms\cite{edmoa74ax}.


\begin{equation}
{\bf D}^{(\alpha,\beta)}(\boldsymbol{\hat{R}}_{\alpha\beta})
= \boldsymbol{R}^{(\mathrm{L}_{\alpha},\mathrm{M}_{\mathrm{l}\alpha})}(\boldsymbol{\hat{R}}_{\alpha\beta})
\otimes
\boldsymbol{R}^{(\mathrm{L}_{\beta},\mathrm{M}_{\mathrm{l}\beta})}(\boldsymbol{\hat{R}}_{\alpha\beta})
  \label{eqn:dmats}
\end{equation}

\noindent Of only computational impact is the fact that the
Hamiltonian matrices, which are all heretofore real valued, become
complex for the most general orientational change.  In terms of
coupling via elements of the transformation matrix, basis function
manifolds with a common set of labels $\mathrm{L}_\alpha$,
$\mathrm{S}_\alpha$, $\mathrm{P}_{\alpha}$, $\mathrm{e}_\alpha$ and
$\mathrm{L}_\beta$, $\mathrm{S}_\beta$, $\mathrm{P}_\beta$,
$\mathrm{e}_\beta$ are intermixed as the labels
$\mathrm{M}_{\mathrm{l}\alpha}^\prime$ and
$\mathrm{M}_{\mathrm{l}\beta}^\prime$ are transformed to reflect their
new values $\mathrm{M}_{\mathrm{l}\alpha}$
$\mathrm{M}_{\mathrm{l}\beta}$.  The sparsity of the total
transformation matrix composed of these submatrices allows
considerable computational efficiency.

\subsection{Diatomic spin uncoupling}

The foundational valence-bond methods are explicitly spin free.
Therefore it is permissible to expand the rotated basis by forming, at
least notionally, $2\mathrm{S}_{\alpha\beta}+1$ space-spin product
functions from each spin-free function with diatomic
$\mathrm{S}_{\alpha\beta}$ for each possible value of diatomic
$\mathrm{M}_{\mathrm{s}\alpha\beta}$.  Furthermore, we are free to
adopt the global polyatomic quantization direction already described
in the space-rotation of orbital angular momentum.  (This is said to
be ``notional'' in the sense that only the Hamiltonian matrices must
be given explicit computational representation, rather than the basis
itself.)  Thus, the matrices in the expanded basis with diatomic
quantum numbers $\mathrm{S}_{\alpha\beta}$ and
$\mathrm{M}_{\mathrm{s}\alpha\beta}$ and atomic labels
$\mathrm{L}_\alpha$, $\mathrm{M}_{\mathrm{l}\alpha}$,
$\mathrm{S}_\alpha$, $\mathrm{P}_{\alpha}$, $\mathrm{e}_\alpha$;
$\mathrm{L}_\beta$, $\mathrm{M}_{\mathrm{l}\beta}$,
$\mathrm{S}_\beta$, $\mathrm{P}_\beta$, $\mathrm{e}_\beta$ can be
constructed by replicating the spin-free matrices along the block
diagonal making use of the orthogonality of the diatomic spin
functions.  This new supermatrix is designated
${\bf H}_{\bf d}^{(\mathrm{S}_{\alpha\beta},
  \mathrm{M}_{\mathrm{s}\alpha\beta})} (\boldsymbol{R}_{\alpha\beta})$.

The diatomic ($\mathrm{S}_{\alpha\beta}$,
$\mathrm{M}_{\mathrm{s}\alpha\beta}$) to atomic
($\mathrm{S}_{\alpha}$, $\mathrm{M}_{\mathrm{s}\alpha}$;
$\mathrm{S}_{\beta}$, $\mathrm{M}_{\mathrm{s}\beta}$) spin-decoupling
coefficients provided by angular-momentum algebra\cite{edmoa74ax}
permit transformation of this basis to one in which the degrees of
freedom of diatomic spin is replaced by atomic counterparts.


\begin{equation}
{\bf H}_{\bf d}^{(\mathrm{M}_{\mathrm{s}\alpha\beta})}
(\boldsymbol{R}_{\alpha\beta}) =
  {\bf T}_{\bf
    d}^{(\mathrm{S}_{\alpha\beta},\mathrm{M}_{\mathrm{s}\alpha\beta},
    \mathrm{S}_{\alpha},\mathrm{M}_{\mathrm{s}\alpha},
    \mathrm{S}_{\beta},\mathrm{M}_{\mathrm{s}\beta})\,t}\cdot
{\bf H}_{\bf d}^{(\mathrm{S}_{\alpha\beta},
  \mathrm{M}_{\mathrm{s}\alpha\beta})}
(\boldsymbol{R}_{\alpha\beta})
  \cdot
  {\bf T}_{\bf d}^{(\mathrm{S}_{\alpha\beta},\mathrm{M}_{\mathrm{s}\alpha\beta},
    \mathrm{S}_{\alpha},\mathrm{M}_{\mathrm{s}\alpha},
    \mathrm{S}_{\beta},\mathrm{M}_{\mathrm{s}\beta})}
  \label{eqn:spuc}
\end{equation}

\noindent This transformation mixes terms with the same
$\mathrm{M}_{\mathrm{s}\alpha\beta}$ but different
$\mathrm{S}_{\alpha\beta}$.  The spectral-product basis which supports
this matrix:


\begin{equation}
  \boldsymbol{\Phi}_{\bf
    d}^{(\alpha,\beta)}(\boldsymbol{i},\boldsymbol{j}) \equiv 
  \{\boldsymbol{\Phi}^{(\alpha)}(\boldsymbol{i})\otimes
  \boldsymbol{\Phi}^{(\beta)}(\boldsymbol{j})\}
  \label{eqn:spbas}
\end{equation}

\noindent has quantum numbers $\mathrm{M}_{\mathrm{s}\alpha\beta}$ (=
$\mathrm{M}_{\mathrm{s}\alpha}$ + $\mathrm{M}_{\mathrm{s}\beta}$) and
atomic labels $\mathrm{L}_\alpha$, $\mathrm{M}_{\mathrm{l}\alpha}$,
$\mathrm{S}_\alpha$, $\mathrm{M}_{\mathrm{s}\alpha}$,
$\mathrm{P}_{\alpha}$, $\mathrm{e}_\alpha$; $\mathrm{L}_\beta$,
$\mathrm{M}_{\mathrm{l}\beta}$, $\mathrm{S}_\beta$,
$\mathrm{M}_{\mathrm{s}\beta}$, $\mathrm{P}_\beta$,
$\mathrm{e}_\beta$.

\subsection{Nature of the orthogonalization reconsidered}

In connection with the orthogonalization of eq.~\ref{eqn:sotrans}, it
was noted that the transformation is dense, that is for a given
$\mathrm{S}_{\alpha\beta}$ and $\lambda_{\alpha\beta}^\prime$, it
extensively mixes functions with different atomic labels.  This would
seem to vitiate the transformations of eqs.~\ref{eqn:rotst}
and~\ref{eqn:spuc} which explicitly rely upon the atomic degrees of
freedom to provide the transformation coefficients.  It might even
seem to necessitate deferring orthogonalization until after those
transformations.  In fact, the nature of symmetric
orthogonalization\cite{lowdp70a,srivv00a} justifies the validity of
the current order.  Not only is symmetric orthogonalization
phase-preserving, but it is also invariant to unitary transformation.
As the space-rotation and spin-uncoupling transformations are unitary,
orthogonalization either before or after leads to the same matrices
${\bf H}_{\bf d}^{(\mathrm{M}_{\mathrm{s}\alpha\beta})}$.  The current
order provides significant computational advantage as it reduces the
size of the ${\bf H}_{\bf d}^{(\alpha,\beta)}(R_{\alpha\beta})$ which
must be stored in the diatomic archive and reduces the size of the
overlap matrices which must be diagonalized to construct the
orthogonalization matrices in eq.~\ref{eqn:somat}.  (This
diagonalization is significantly more expensive than the
transformations of eqs.~\ref{eqn:rotst} and~\ref{eqn:spuc} which must,
nevertheless, be performed for each pair orientation in the polyatomic
and repeated for any change of geometry.)

One other property of symmetric orthogonalization is important for the
current discussion.  It can be shown to produce orthogonal basis
vectors which are collectively closest (in a least-squares sense) to
the original basis\cite{lowdp70a,srivv00a}.  It is of course one thing
to determine that a particular orthogonal set is the closest possible
(in some sense) to the original basis and yet another to conclude that
the two bases are particularly close in an absolute sense.  Thus, the
underlying functions of the rotated and spin-uncoupled orthonormal
basis can be given only approximate or provisional atomic labels
deriving from their ancestry in the valence-bond functions
(eq.~\ref{eqn:vbbas}) and strictly valid only at infinite separation.
To recap, these atomic labels include $\mathrm{L}$,
$\mathrm{M}_{\mathrm{l}}$, $\mathrm{S}$, $\mathrm{M}_{\mathrm{s}}$,
$\mathrm{P}$ and $\mathrm{e}$ for each atom.  The symmetries observed
in the polyatomic Spectral Theory Hamiltonian matrix will be shown in
a later section to cast further light on this issue.

\subsection{Summary of diatomic basis dimension and ordering}

As an aid to understanding the nature of the various bases and
transformations involved in the preceding steps it is illuminating to
consider as a simple example the CH diatomic molecule.  If the carbon
atom is regarded as having a frozen 1s core and active 2s and 2p
electrons, an atomic CI calculation can be performed using all the
$\mathrm{2s^2\,2p^2,}$ $\mathrm{2s^1\,2p^3,}$ and
$\mathrm{2s^0\,2p^4}$ configurations.  (Restricted Hartree-Fock
calculations for the atomic ground state can be used to provide
one-electron orbitals for the configurations.)  There are then 36
solutions of eq.~\ref{eqn:atsol}, each with a unique set of values of
S, L, $\mathrm{M_l}$, P, and e.  In the usual atomic term symbols
(with e, if required, appended in parentheses), the manifolds
represented are $\mathrm{^3P_g(1)}$, $\mathrm{^1D_g(1)}$,
$\mathrm{^1S_g(1)}$, $\mathrm{^5S_u}$, $\mathrm{^3D_u}$,
$\mathrm{^3P_u}$, $\mathrm{^1D_u}$, $\mathrm{^3S_u}$,
$\mathrm{^1P_u}$, $\mathrm{^3P_g(2)}$, $\mathrm{^1D_g(2)}$, and
$\mathrm{^1S_g(2)}$.  If hydrogen has 3 s-orbitals and 1 set of
p-orbitals, the total dimension is 6 and there are 4 manifolds,
$\mathrm{^2S_g}(1)$, $\mathrm{^2S_g}(2)$, $\mathrm{^2P_u}$, and
$\mathrm{^2S_g}(3)$.

Diatomic CI calculations can then be performed using as configurations
all combinations of the carbon and hydrogen orbital occupations listed
above.  The Hamiltonian and overlap matrices of eq.~\ref{eqn:diytsolv}
then have a dimension of 322 with 10 symmetry subblocks identified
with: {$^2\Sigma$, $^2\Pi$, $^2\Delta$, $^2\Phi$, $^4\Sigma$, $^4\Pi$,
  $^4\Delta$, $^4\Phi$, $^6\Sigma$, and $^6\Pi$}.  The solutions of
eq.~\ref{eqn:diytsolu} have the same multiplicity and symmetry
character.  The valence-bond basis of eq.~\ref{eqn:vbbas} which
underlies the Hamiltonian (and overlap) matrices of
eq.~\ref{eqn:yttovb} have the same dimension and diatomic quantum
numbers.  Finally the archived Hamiltonian matrices
(eq.~\ref{eqn:sotrans}) in the antisymmetrized, orthogonalized-product
basis have the same arrangement with approximate atomic labels.  As an
example of the denseness of the orthogonalization transformation, at
small separation the orthogonal basis state of $^2\Sigma$ symmetry which has a
nominal label of C: $\mathrm{^1D^{m_l=-1}_g}(1)\otimes$ H:
$\mathrm{^2P^{m_l=+1}_u}$ has largest contribution from the original
basis state of the same label, but significant contributions from the
remaining 59 states which arise from the allowable couplings of
singlet and triplet carbon states with doublet hydrogen states.
Thus, properly speaking, the nominal label should actually be replaced
by ``the orthogonalized, antisymmetric, diatomic basis state closest
to the product of carbon in the $\mathrm{^1D^{m_l=-1}_g}(1)$ state and
hydrogen in the $\mathrm{^2P^{m_l=+1}_u}$ state.''  The nominal label
will often be used in the interests of brevity.

Under rotation (eq.~\ref{eqn:rotst}) these 322 basis functions produce
non-zero Hamiltonian elements distributed among 3 subblocks separated
only by diatomic total spin ($\mathrm{S}_{\alpha\beta}$ = 1/2, 3/2,
and 5/2).

The $\mathrm{M}_{\mathrm{s}\alpha\beta}$-expanded matrix
${\bf H}_{\bf d}^{(\mathrm{S}_{\alpha\beta},
  \mathrm{M}_{\mathrm{s}\alpha\beta})} (\boldsymbol{R}_{\alpha\beta})$
on the right-hand side of eq.~\ref{eqn:spuc} has a dimension of 840
with nonzero values spread among 12 subblocks distinguishable by
$\mathrm{S}_{\alpha\beta}$ and $\mathrm{M}_{s\alpha\beta}$.
Spin-coupling to the left-hand side of eq.~\ref{eqn:spuc} rearranges
the supermatrix of dimension 840 into 6 subblocks consistent with the
allowable values of diatomic spin projection,
$\mathrm{M}_{s\alpha\beta}$ = $\pm 1/2,$ $\pm 3/2$, and $\pm 5/2$.

The total dimension of the diatomic basis here is the same as that
which results from the direct product of the 70 energy-ordered carbon
states and 12 hydrogen states:


\begin{equation}
\mathrm{C:}\> \{\mathrm{^3P_g}(1), \mathrm{^1D_g}(1), \mathrm{^1S_g(1)},
\mathrm{^5S_u}, \mathrm{^3D_u}, \mathrm{^3P_u},
\mathrm{^1D_u}, \mathrm{^3S_u}, \mathrm{^1P_u},
\mathrm{^3P_g}(2), \mathrm{^1D_g}(2), \mathrm{^1S_g(2)}\}
\label{eqn:cbasex}
\end{equation}


\begin{equation}
\mathrm{H:}\> \{\mathrm{^2S_g}(1), \mathrm{^2S_g}(2), \mathrm{^2P_u}, 
  \mathrm{^2S_g}(3)\}
\label{eqn:hbasex}
\end{equation}

\noindent The state counting has included all the spin- and
space-angular-momentum degrees of freedom, parity, and the multiple
states of the same symmetry in the atomic valence spaces.  This is
referred to in this work as the ``multiplicity'' of the ``atomic
spectral basis'' and has special significance as, together with those
of the other atoms, it determines the size of the polyatomic matrices
which must ultimately be diagonalized.

The exclusive use of
neutral-atom states at this point might seem to be inadequate,
especially in comparison with other methods which build up polyatomic
molecules from fragment states.  As the spectral basis is increased in
size, the excited states above atomic ionization collectively
represent the effect of charged states embedded in the continuum (see
ref.~\citenum{millj16a}, sect. 4).  Explicit addition of charged
states to this description would lead to linear dependence in the
underlying Spectral Theory basis.

Independent of the degree to which the final diatomic basis functions
of eq.~\ref{eqn:spbas} can be assigned definitive atomic-state
labels (because of mixing in orthogonalization) it is essential to
understand that each of the functions is rigorously antisymmetric in
the exchange of all of the electrons in the diatom.  The inclusion in
finite-basis atomic versions of the Spectral Theory of nonphysical
states which are not totally antisymmetric with respect to electron
exchange and the need to project them out of a relatively expanded
basis to isolate the physical solutions has been described in previous
publications on the Spectral Theory.  These non-Pauli contaminants which belong to other
irreducible representations of the symmetric group\cite{hamem64ax} are
totally absent from the present approach.

Finally, it is worth commenting on the necessity of carefully tracking
the order of the diatomic basis functions, in particular the rows and
columns of the matrices to which they correspond.  The default order
is referred to as ``odometer ordering'' and runs through the degrees
of freedom of the second atom (hydrogen here) before the first
(carbon), exhausting $\mathrm{M_s}$ and then $\mathrm{M_l}$ labels
(lowest to highest) for each of the atomic states in moving in order
(left to right) through the atomic state manifolds listed in
eqs.~\ref{eqn:cbasex} and~\ref{eqn:hbasex}.  In the course of the
transformations itemized in this section, the changing quantum numbers
of the basis can lead to block-diagonality which may be masked or
manifested with different provisional orderings for the different
steps.  Computational efficiency can be dramatically enhanced by
temporarily changing the order to minimize the size of the sub-blocks
which need be manipulated.  Correctly reversing this reorganization
between steps is essential.

\subsection{Polyatomic solutions from pair matrices}

As described in previous publications, the two-body nature of the
Coulomb interactions allows the polyatomic, Born-Oppenheimer,
quantum-chemical Hamiltonian to be exactly rearranged to an
atomic-pairwise form:


\begin{equation}
  H(\boldsymbol{r}:\boldsymbol{R}) =
  \sum_{\alpha=1}^{\mathrm{N}-1}\sum_{\beta = \alpha+1}^\mathrm{N}
  \Bigl\{
    H^{(\alpha,\beta)}(\boldsymbol{i,j}:\boldsymbol{R}_{\alpha\beta})
    -\frac{\mathrm{N}-2}{\mathrm{N}-1}\{
    H^{(\alpha)}(\boldsymbol{i}) + H^{(\beta)}(\boldsymbol{j})
    \}
    \Bigr\}
  \label{eqn:polyham}
\end{equation}

\noindent The second term in the summation removes the overcounting of
the intra-atom interactions in the sum over the pair terms.  The
assignment of particular electrons to particular nucleii is not
problematic when applied in the context of the antisymmetrized product
basis of the Spectral Theory.  (See especially the discussion of eq. 4
in ref.~\citenum{langp18a}.)

Taking advantage of the pair separation in the summation of the
operators in the previous equation, the polyatomic Hamiltonian matrix
can be composed from diatomic matrices as:


\begin{equation}
  {\bf H}(\boldsymbol{R}) =
  \sum_{\alpha=1}^{\mathrm{N}-1}\sum_{\beta = \alpha+1}^\mathrm{N}
  \Bigl\{
    {\bf H}_{\bf d}^{(\alpha,\beta)}(\boldsymbol{R}_{\alpha\beta})
    -\frac{\mathrm{N}-2}{\mathrm{N}-1}\{
    {\bf H}_{\bf d}^{(\alpha,\beta)}({\bf R}_{\alpha\beta}\rightarrow \infty)
    \}    
    \Bigr\}
  \label{eqn:polydimat}
\end{equation}

\noindent The atomic-energy-overcounting correction has been
implemented with a pair-matrix in which the atoms have been removed to
infinite separation.  (ref.~\citenum{langp18a} refers to this equation
as defining the ``factored antisymmetric atomic-pair form in the
finite subspace spectral-product basis.'') The polyatomic
eigensolutions in the finite, spectral-product basis are then obtained
by diagonalization of this matrix.

One of the essential demonstrations of a previous report on the
Spectral Theory\cite{langp08a} was that a finite,
pairwise-antisymmetrized basis, such as underlies the
${\bf H}_{\bf d}^{(\alpha,\beta)}(\boldsymbol{R}_{\alpha\beta})$
matrices, directly provides a basis for polyatomic solutions which are
pure Pauli states.  Unphysical diatomic states need not be eliminated
(as discussed previously) and this physical purity propagates to the
polyatomic solutions.  Conversely, atom-based variants of the Spectral
Theory short of the closure limit seem to involve the possibility of
contamination of molecular solutions with modest or small (in absolute
weight) nonphysical contributions with
unknown quantitative effect upon calculated polyatomic
energies. Those earlier development are founded directly upon the
odometer-ordered atomic product basis:


\begin{equation}
  {\boldsymbol{\Phi}}(\boldsymbol{r}:\boldsymbol{R}) =
  \{
  \boldsymbol{\Phi}^{(1)}(\boldsymbol{1})\otimes
  \boldsymbol{\Phi}^{(2)}(\boldsymbol{2})\otimes ... \otimes
  \boldsymbol{\Phi}^{(\mathrm{N})}(\boldsymbol{n})
  \}_O
  \label{eqn:stpolybas}
\end{equation}

In the current approach this potential problem is prevented, but
perhaps at the cost of ambiguity in the atomic character of the states
in the finite-basis polyatomic product basis.  In contrast to the
basis of eq.~\ref{eqn:stpolybas}, it is conceivable that states with
the same nominal labels for an atom which is part of two different
pairs in an aggregate may actually be slightly different.  This might
stem from the divergence of the original and orthogonal basis
(eq.~\ref{eqn:sotrans}) as distance decreases.  Although this
potential ambiguity does not complicate the numerical procedures involved in
matrix construction and eigensolution it may be problematic
for interpretation of the solutions, particularly for any atomic or
diatomic partitioning schemes which rely, unless only in an
approximate sense, upon the atomic character of the polyatomic basis
functions.

Finally, an important observation provides a polyatomic perspective
upon the ``least-squares-closeness'' of the original and orthogonal
diatomic bases and the symmetry-preserving nature of symmetric
orthogonalization.  All polyatomic matrices constructed via
eq.~\ref{eqn:polydimat} are block diagonal in the sum of the
(approximate or nominal) atomic $\mathrm{M_s}$ labels.  Additionally,
linear molecules oriented along the quantization axis have polyatomic
matrices which are block diagonal in the sum of atomic $\mathrm{M_l}$
labels.  For molecules with an axis of symmetry and belonging to some
of the simpler Abelian point groups, symmetry adaptation based on the
$\mathrm{M_l}$ values of symmetry-unique atoms and yielding matrices
blocked by irreducible representation has even been possible.
Apparently the transformation of eq.~\ref{eqn:sotrans}, although dense
in mixing states of different atomic labels, not only provides
functions with nominal atomic labels close in a least-squares sense to
functions with absolute atomic character, but, even through the
summations of eq.~\ref{eqn:polydimat}, preserves fundamental atomic
symmetries.

\subsection{Summary of polyatomic basis ordering and dimension}

The basis underlying the Hamiltonian matrix of eq.~\ref{eqn:polydimat}
is odometer-ordered as eq.~\ref{eqn:stpolybas}, in analogy with the
procedure already described with the diatomic example, once the atomic
order has been specified.

Consider the $\mathrm{CH_2}$ molecule in the carbon and hydrogen bases
previously used in the diatomic example (eqs.~\ref{eqn:cbasex}
and~\ref{eqn:hbasex}).  Before block-diagonal size reductions due to
space and spin symmetries are taken into account, the total basis
dimension is 10,800, just the product of the number of states
describing each atom.

For a general molecule composed of a total of $\mathrm{N}$ atoms
with $\mathrm{N_k}$ atoms of element type $\mathrm{k,}$ each atomic
element of which has an atomic state multiplicity of $\mathrm{n_i}$,
the total dimension of the molecular basis (eq.  ~\ref{eqn:stpolybas})
is:

\begin{equation}
  \mathrm{D} = \prod_\mathrm{i} \mathrm{n_i^{N_i}} 
  \label{eqn:stbasdim}
\end{equation}

\noindent Thus, the size of Spectral Theory Hamiltonian scales
polynomially in the atomic state multiplicity and exponentially in the
number of atoms.

This has the same dimension as an analogous atom-centered valence-bond
calculation in the polyatomic Young-tableau basis.  Modest relative
computational advantages of the Spectral Theory may be attributed to a
need to solve only the overlap-free, ordinary-eigenvalue problem and
the relative sparseness of the Spectral Theory Hamiltonian matrix.
More profound advantages result from the total absence of two-electron
integrals over more than two centers and the ``once for all''
calculation of the diatomic archive matrices.  In this report the
polynomial scaling of the polyatomic basis with the number of atomic
states will motivate efforts to economize by
reducing the number of states needed to accurately describe each atom
and by injecting more accurate diatomic curves into treatments with
given atomic dimension.  These two procedures are more easy to
understand in the context of the full basis results and so their
description will be deferred.  Subsequent publications will address
approximations which enable breaking of the polynomial scaling itself.

\section{Full-basis results}

\subsection{Computational issues}

Because the archived diatomic matrices on the left-hand side of
eq.~\ref{eqn:sotrans} need only be calculated once, otherwise
expensive one-electron Slater bases can be more easily accommodated.  Energy-optimized, valence
bases (VB1,VB2,VB3)\cite{emaai03a} and the SMILES Slater integral
package\cite{ricoj01a} are used to calculate the one- and two-electron
integrals (over a maximum of two atomic centers).  The valence-bond
CRUNCH suite\cite{CRUNCH} is then used for the remaining atomic and
diatomic steps which involve Young tableaux.  Code specially written
for the Spectral Theory performs the remaining steps.  Most of the
computational expense involves diagonalization, either to construct
the orthogonalization matrices of eq.~\ref{eqn:somat} or, especially,
to find solutions of the polyatomic matrix of eq.~\ref{eqn:polydimat}.
If the matrices can be held in memory, the LAPACK libraries are used.
For larger problems methods in which the matrices are partitioned,
such as one already in CRUNCH\cite{gallg82c} and the
PRIMME\cite{stata10a,PRIMME} libraries proved useful.  On-the-fly
calculation of the polyatomic matrix elements on demand with only the
diatomic matrices of the right-hand-side of eq.~\ref{eqn:polydimat}
held in memory was sufficiently fast to be used with iterative matrix
diagonalization.  Geometry optimization was performed using a simplex
optimizer.\cite{presw07ax} The PETSC libraries\cite{PETSc} were used
to provide interlanguage communication between Fortran and C modules.
The module to rotate the phase-consistent diatomic matrices was
adapted from code provided by the authors of the package which
calculates the Slater integrals, as it observes the same
angular-momentum conventions.

\subsection{Atomic results}

The one-electron orbitals from which the Young-tableaux configurations
are constructed are the SCF solutions, both valence and virtual, for
the ground state of the atoms in the three VB bases.
Table~\ref{tbl:hspect} shows the eigenvalues of the CI problem of for
the hydrogen atom where the single electron can be excited to any of
the virtual orbitals.  (If the single electron is not allowed to be
excited, then, of course, there is a single $\mathrm{^2S_g}$ state
with a multiplicity of 2.)  The difference in absolute energies (not
displayed in the table) of the ground state between the three bases
and with or without excitation are insignificant.  The creation energy
of the bare proton via ionization is included for reference.  The
indicated multiplicities of the atomic spectrum are calculated to
include spin ($\mathrm{M_s}$) degrees of freedom and these
multiplicities are the number of atomic states which appear in the
product over atoms which determine the dimension of the polyatomic
problem (eq.~\ref{eqn:stbasdim}).  Even in the largest basis, only the
functions with n=2 are directly represented with the rest of the virtual space
composed of increasing numbers of post-threshold states.


\begin{table}
  \caption{Full-excitation hydrogen atom spectra (eV)}
  \label{tbl:hspect}
  \begin{tabular}{rrrr}
    \hline
    Basis            &   VB1 &  VB2 &   VB3 \\
    \hline
    Multiplicity\textsuperscript{\emph{a}} &   12 &    30 &   62 \\
    \hline
    State            &      &       &      \\
    \hline
    $\mathrm{^2S_g}$ &   0.0 &   0.0 &   0.0 \\
    $\mathrm{^2P_u}$ &       &  10.2 &  10.2 \\
    $\mathrm{^2S_g}$ &       &  12.2 &  10.2 \\
    $\mathrm{[H^+]}$ &  13.6 &  13.6 &  13.6 \\
    $\mathrm{^2S_g}$ &  16.2 &  16.1 &  15.0 \\
    $\mathrm{^2P_u}$ &  23.3 &  48.5 &  16.3 \\
    $\mathrm{^2D_g}$ &       &  58.9 &  27.3 \\
    $\mathrm{^2S_g}$ & 125.1 &  71.8 &  42.9 \\
    $\mathrm{^2P_u}$ &       &       &  56.6 \\
    $\mathrm{^2D_g}$ &       &       & 103.4 \\
    $\mathrm{^2F_u}$ &       &       & 110.6 \\
    $\mathrm{^2S_g}$ &       &       & 244.7 \\
    \hline
  \end{tabular}

  \textsuperscript{\emph{a}} See discussion of example,
  eqs.~\ref{eqn:cbasex} and ~\ref{eqn:hbasex}
\end{table}

Table~\ref{tbl:cspect} shows some corresponding results for the carbon
atom with a frozen 1s core.  The data in the first column are
calculated using the VB3 basis and what are referred to in the
Spectral-Theory context as ``no excitation'' or ``valence-excitation''
calculations.  In particular excitation outside of the 2s and 2p
orbital space is forbidden.  (That is, this is full CI within a
restricted orbital space).  As the direct product of these atomic
spaces are used for the diatomic calculations, this is the
atom-localized equivalent of the reference spaces employed in
conventional MOCI calculations.  The energies of the valence states
(along with the ground-state of the singly charged cation, for
reference) are given in the table with the prominence of the
valence-excited states above the ionization threshold especially to
be noted.  (Corresponding valence-excitation results for the other two
bases are not shown as they have an absolute difference from the
ground state of less than 0.01 eV and term energies which are
identical to within the rounding in the table.)  The next three
columns show results for the three bases in which single excitation from the
reference to the non-reference orbitals is allowed.  The absolute ground-state energies (not given) differ among
themselves by less than 0.1 eV and are slightly more than 1 eV below
the no-excitation results.  Approximate assignments of the
valence-excited states have been attempted, with the understanding
that these are at best tentative, as spectral density is generally
distributed among several states of the same symmetry.  The energies
of the states represented by the difference between 70 (the
no-excitation multiplicity) and the indicated basis multiplicity have
not been displayed but have an increasing density of states up to the
maximum state energy given in the last line.  The next two columns
(``A'' and ``B'') are included for ease of later comparison, but
discussion will be deferred until the next section.  The final column
contains the experimental results.


\begin{table}
  \caption{Valence portion of the frozen-core atomic spectral bases for carbon atom (eV)}
  \label{tbl:cspect}
  \begin{tabular}{lrrrrrrr}
    \hline
    Basis  & VB3*\textsuperscript{\emph{a}}
    &  VB1 &  VB2 &   VB3 &A\textsuperscript{\emph{b}}&B\textsuperscript{\emph{b}}&Exp.\textsuperscript{\emph{c}}\\
    \hline
    Multiplicity\textsuperscript{\emph{d}}&  70 &  1638 &  3430 & 6230 &  238 & 1090&     \\
    \hline
    State             &      &       &      &      &      &      &     \\
    \hline
    $\mathrm{^3P_g(1)}$&  0.0 &  0.0 &  0.0 &  0.0 &  0.0 &  0.0 &   0.0\\
    $\mathrm{^1D_g(1)}$&  1.6 &  1.6 &  1.5 &  1.5 &  1.6 &  1.6 &   1.3\\
    $\mathrm{^1S_g(1)}$&  2.7 &  3.6 &  3.6 &  3.6 &  2.6 &  3.1 &   2.7\\
    $\mathrm{^5S_u}$   &  3.0 &  4.0 &  4.0 &  4.0 &  3.0 &  3.5 &   4.2\\
    $\mathrm{^3D_u}$   &  8.5 &  8.4 &  8.2 &  8.2 &  8.1 &  8.4 &   7.9\\
    $\mathrm{^3P_u}$   & 10.1 & 10.4 &  9.5 &  8.5 &  9.7 &  8.7 &   9.3\\
    $\mathrm{[C^+]}$   & 11.4 & 10.6 & 10.6 & 10.6 & 11.4 & 11.9 &  11.3\\
    $\mathrm{^1D_u}$   & 14.7 & 13.5 & 12.9 & 12.8 & 13.5 & 13.8 &  12.9 \\
    $\mathrm{^3S_u}$   & 15.5 & 14.3 & 13.6 & 13.6 & 15.5 & 13.7 &  13.1 \\
    $\mathrm{^1P_u}$   & 16.3 & 15.6 & 15.7 & 14.6 & 15.9 & 15.0 &  14.3 \\
    $\mathrm{^3P_g(2)}$& 21.2 & 17.1 & 16.5 & 14.3 & 19.5 & 18.3 &   \\
    $\mathrm{^1D_g(2)}$& 22.8 & 19.5 & 20.5 & 18.7 & 20.8 & 18.4 &   \\
    $\mathrm{^1S_g(2)}$& 26.5 & 22.5 & 23.3 & 21.3 & 22.8 & 20.7 &   \\
    $\mathrm{[max]}\textsuperscript{\emph{e}}$   &  -- & 1432.3&1735.1&1896.9& 26.6 & 36.2 &   \\
    \hline
  \end{tabular}
  
  \textsuperscript{\emph{a}} Valence excitation only (see text)
  \textsuperscript{\emph{b}} Results included for comparison, not
  discussed until later section
  \textsuperscript{\emph{c}} Ref.~\citenum{NIST}
  \textsuperscript{\emph{d}} See discussion of example,
  eqs.~\ref{eqn:cbasex} and ~\ref{eqn:hbasex}
  \textsuperscript{\emph{e}} Energy of highest state
\end{table}

\subsection{Diatomic results}

\begin{figure}
  \includegraphics[width=1.0\textwidth]{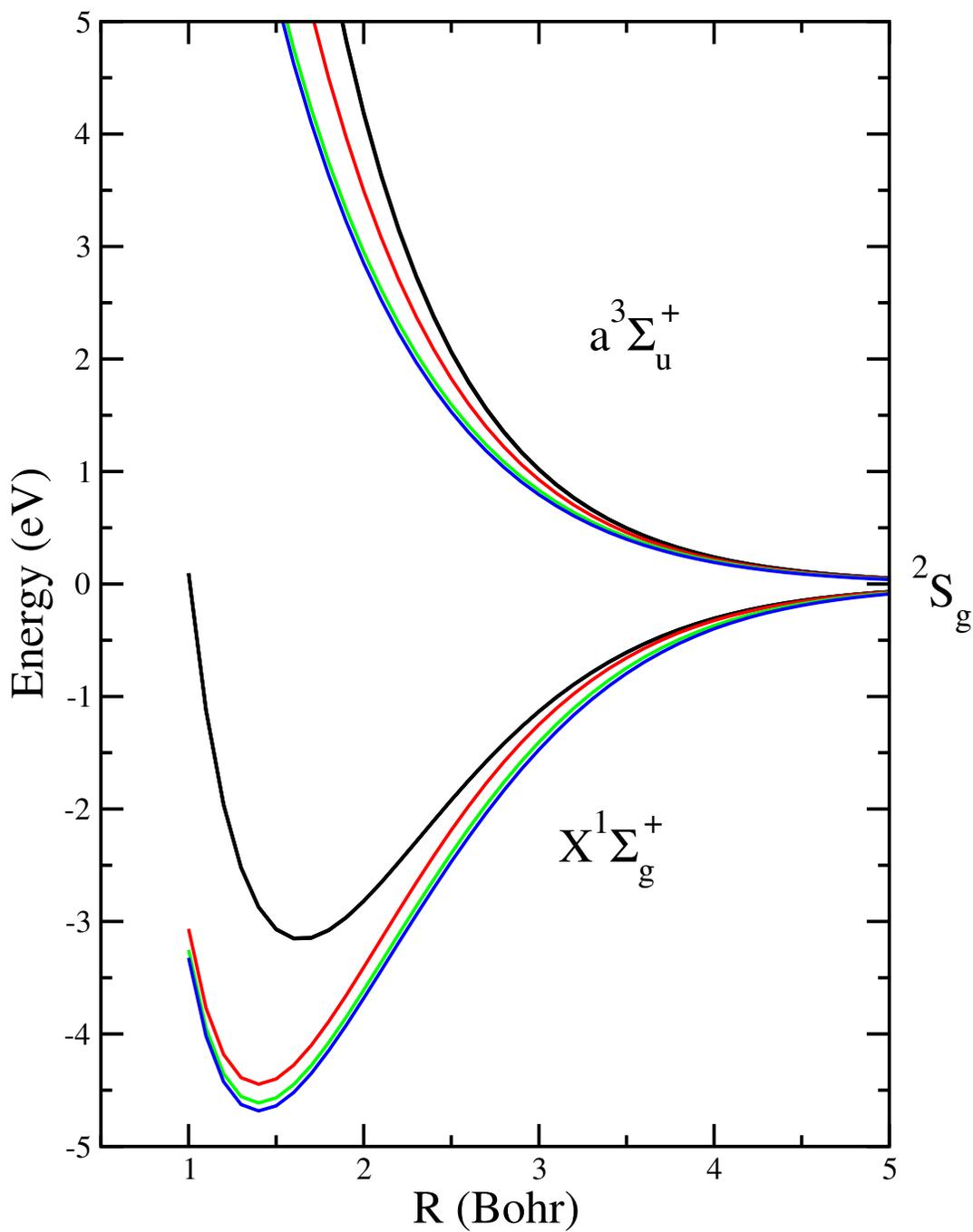}
  \caption{Diatomic-hydrogen interaction energies (lowest two
    states).  Black, 1s-only solutions; Red, VB1-single-excit.;
    Green, VB2-single-excit.; Blue, VB3-single-excit.  Asymptotic
    atomic state on the right.}
  \label{fgr:h2PES}
\end{figure}

Figure~\ref{fgr:h2PES} shows diatomic hydrogen interaction energies
(calculated as in eqs.~\ref{eqn:diytsolv} or~\ref{eqn:distsol}) for
the lowest singlet and triplet states in the same bases.  Three black
lines (indistinguishable) give the no-excitation (Heitler-London-type)
VB1, VB2, and VB3 results and red, green, and blue lines are for
single-excitation VB1, VB2, and VB3, respectively.  The asymptotic
atomic state ({\it c.f.} tab.~\ref{tbl:hspect}) is included on the
right side.


\begin{figure}
  \includegraphics[width=1.0\textwidth]{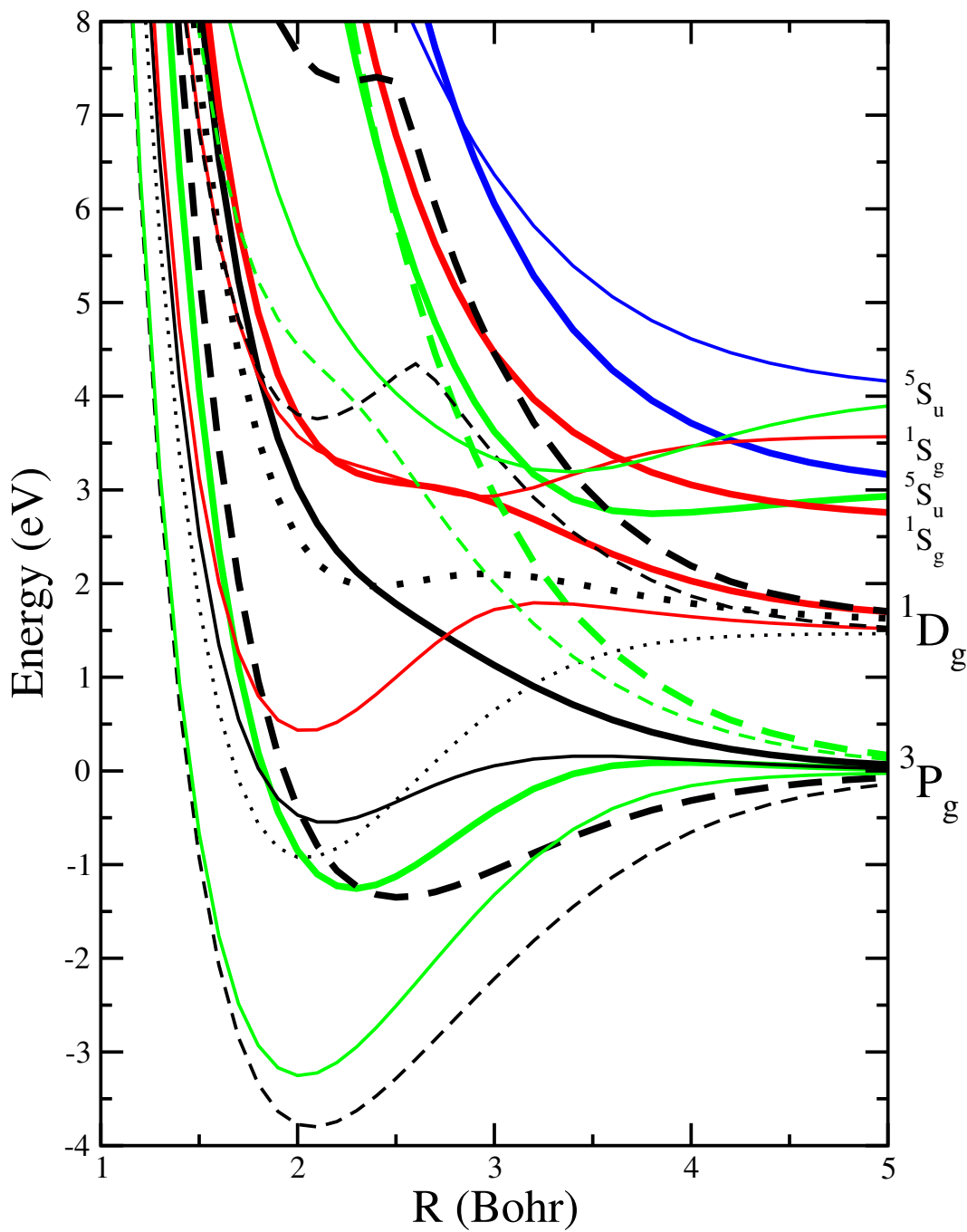}
  \caption{Diatomic CH interaction energies dissociating to the lowest
    four asymptotes in the VB3 basis.  Thicker curves-no excitation;
    thinner curves-single excitation.  Solid lines-$\Sigma$; dashed
    lines-$\Pi$, dotted lines-$\Delta$.  Black-doublets (including
    $^2\Sigma^-$); green-quartets; blue-sextets; exception:
    red-$^2\Sigma^+$. Asymptotic carbon asymptotes paired with
    $\mathrm{^2S_g}$ hydrogen are noted on the right; energy differences in the
    higher asymptotes can be noted in accord with the values in
    table~\ref{tbl:cspect}.  }
  \label{fgr:chPES}
\end{figure}

Figure~\ref{fgr:chPES} shows some similar results for CH.  The thicker
lines (consistently less bound) are the VB3 no-excitation results.
The thinner lines are for the VB3 basis with up to single excitation
on each atom.  It should be noted that these (and later reported
polyatomic energies) are given in terms of a total, uncorrected
atomization energy, an especially stringent test of theoretical
methods, particular when used for homolytic bond
cleavage.\cite{dutta03a}


The results calculated by more conventional orbital-based CI
techniques in eq.~\ref{eqn:diytsolv} and those in the atomic state
basis of eq.~\ref{eqn:distsol} are identical.  Therefore comparison of
the atomic state bases (tabs.~\ref{tbl:hspect} and~\ref{tbl:cspect})
and the degree of diatomic binding supported by each, illustrate the
perspective which the Spectral Theory provides on the
importance of excited atomic states, even those above the ionization
threshold, for molecular binding.  This is true in the case of
hydrogen where there is a one-to-one correspondence between orbitals
and states as well as in carbon with its multi-electron character.

In moving from diatomic to polyatomic calculations, although all
states are likely important to some degree, it might be supposed that
aggregate binding may be mostly attributable to the most tightly bound
diatomic states.  Table~\ref{tbl:CHnBE} contains the energy minima for
the low-lying diatomic curves for side-by-side comparison.  (As in the
previous table, the two additional columns in this and the following
table will be discussed subsequently.)  In lieu of experimental values
the results of high-level calculations\cite{kolow68a,kalea99b} are
given for comparison.

It should be appreciated that although the underlying CI-based methods
(eq. ~\ref{eqn:diytsolv}) are both size-consistent and variational,
the variational discrepancies of a diatom at its equilibrium
separation and at infinite separation may be different in magnitude.
Thus the diatoms may be calculated to be over or underbound in a given
finite basis.


\begin{table}
  \caption{Diatomic binding energies (eV)}
  \label{tbl:X2BE}
  \begin{tabular}{lrrrrrrr}
    \hline
    Basis                     &VB3*\textsuperscript{\emph{a}}
                              &VB1 &VB2 &VB3&C\textsuperscript{\emph{b}}&D\textsuperscript{\emph{b}} &Ref.\\
    \hline
    CH Multiplicity\textsuperscript{\emph{c}}&140&19,656&102,900&386,260&140&2856&\\
    \hline
    Molec./State               &    &    &    &    &    &    &     \\
    \hline
    $\mathrm{H_2\,^1\Sigma^+_g}$& 3.2& 4.4& 4.6& 4.7 & 4.7& 4.7&4.7\textsuperscript{\emph{d}} \\
    $\mathrm{CH\,^2\Pi}$       & 1.3& 3.1& 3.6& 3.8 & 3.8& 3.8&3.6\textsuperscript{\emph{e}} \\
    $\mathrm{CH\,^4\Sigma^-}$  & 1.3& 2.6& 3.0& 3.3 & 3.3& 3.3&2.9\textsuperscript{\emph{e}} \\
    \hline
  \end{tabular}

  \textsuperscript{\emph{a}} Valence excitation only (see text)
  \textsuperscript{\emph{b}} Results included for comparison, not
  discussed until later section
  \textsuperscript{\emph{c}} See eq.~\ref{eqn:stbasdim}
  \textsuperscript{\emph{d}} Ref.~\citenum{kolow68a}
  \textsuperscript{\emph{e}} Large basis MRCI, ref.~\citenum{kalea99b}
\end{table}

\subsection{Polyatomic results}

\begin{figure}
  \includegraphics[width=1.0\textwidth]{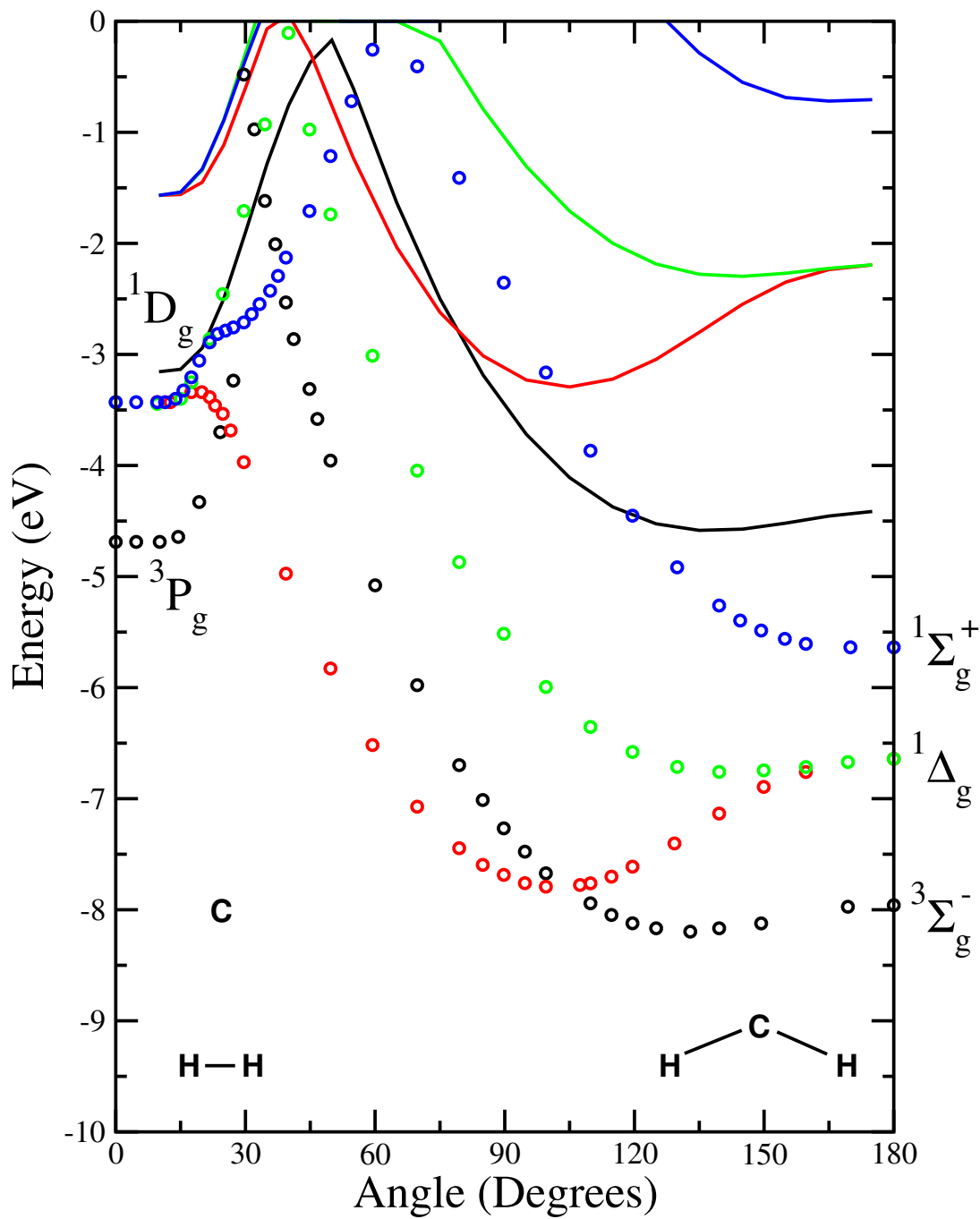}
  \caption{Optimized (see text for procedure and left and right
    asymptotic assignments) $\mathrm{CH_2}$ ($\mathrm{C_{2v}}$)
    total binding energies for fixed angles.  Black-X$\mathrm{^3B_1}$;
    red-a$\mathrm{^1A_1}$;
    green-b$\mathrm{^1B_1}$; blue-c$\mathrm{^1A_1}$.  Lines-Spectral Theory using the
    VB3 bases and only valence-excitation; circles-large-basis MRCI
    results, ref.~\citenum{kalea04a}.}
  \label{fgr:ch2PES}
\end{figure}

Figure~\ref{fgr:ch2PES} displays binding energies to total atomization
calculated from the polyatomic matrix of eq.~\ref{eqn:polydimat} for
the four lowest states of the methylene molecule using the
$\mathrm{H_2}$ and CH diatomic matrices calculated in the VB3 bases
with limitation to valence-excitation.  This system was an important
touchstone in the demonstration of the qualitative and quantitative
partnership of quantum chemistry with spectroscopy in the
characterization of small molecules\cite{goddw85a,schah86a}.  The
optimization proceeds as follows.  For a fixed HCH bond angle
(ordinate) in the ``V-shaped'' $\mathrm{C_{2v}}$ molecule, the CH bond
length is optimized and the binding energy (ordinate) is calculated.
Thus, on the left, a ground-state diatomic hydrogen molecule is
loosely associated with the indicated state of the carbon atom while
on the right, the molecule takes on a symmetric collinear geometry and
the states correlate with those of labeled electronic symmetry.  The
circles give high-level theoretical results (MRCI,
ref.~\citenum{kalea04a}) for comparison.  The evident underbinding in
the Spectral Theory results is not surprising given the underbinding
in the corresponding pair results in table~\ref{tbl:X2BE} (first
column) and is propagated to the polyatomic Hamiltonian matrix by the
finite- and infinite-separation diatomic matrices.  It is telling, however, that the
subtle, qualitative angle-dependence is reproduced.


For side-by-side comparison table~\ref{tbl:CHnBE} contains optimized
(in all degrees of freedom) binding energies for a number of molecules
containing a single carbon and a varying number of hydrogens using
bases already discussed.  Although the
valence-excitation VB3 calculation is relatively primitive it is worth
noting that the $\mathrm{CH_5}$ molecule (VB3* multiplicity 2,240) is
unbound with respect to methane and hydrogen atom, at least initially
alleviating concerns that a polyatomic theory build only from diatomic
information may not observe chemical valence.  (Because wave-function
symmetry adaptation has not been implemented for all point groups
$\mathrm{CH_3}$, $\mathrm{CH_4}$, and $\mathrm{CH_5}$ are calculated
with $\mathrm{C_{2v}}$ electronic symmetry but with a higher symmetry
nuclear geometry, as indicated in parentheses for each.)



\begin{table}
  \caption{Polyatomic binding energies (eV)}
  \label{tbl:CHnBE}
  \begin{tabular}{lrrrrrr}
    \hline
    Basis             & VB3*\textsuperscript{\emph{a}}& VB1 & C\textsuperscript{\emph{b}}&D\textsuperscript{\emph{b}}&Ref.\\
    \hline
    $\mathrm{CH_2}$ Multiplicity\textsuperscript{\emph{c}}&280&235,872&280&34,272&\\
    \hline
    Molec./State              &     &     &    &      &     \\
    \hline
    $\mathrm{CH_2\,X: ^3B_1}$ & 4.6& 7.5 & 9.3& 8.7 &8.2\textsuperscript{\emph{d}} \\
    $\mathrm{CH_2\,a: ^1A_1}$ & 3.3& 7.3 & 8.7& 7.8 &7.8\textsuperscript{\emph{d}} \\
    $\mathrm{CH_2\,b: ^1B_1}$ & 2.3&    & 8.4& 7.4&6.8\textsuperscript{\emph{d}} \\
    $\mathrm{CH_2\,c: ^1A_1}$ & 0.7&    & 7.3&  5.8&5.6\textsuperscript{\emph{d}} \\
    $\mathrm{CH_3\,X(D_{3h})}$& 6.5&     &14.9& 13.4&12.6\textsuperscript{\emph{e}} \\
    $\mathrm{CH_4\,X(T_d)}  $&10.4&     &19.6&    &17.1\textsuperscript{\emph{f}} \\
    $\mathrm{CH_5\,X(D_{3h})}$& 8.2&     &17.1&     &    \\
    \hline
  \end{tabular}

  \textsuperscript{\emph{a}} Valence excitation only
  \textsuperscript{\emph{b}} Results included for comparison, not
  discussed until later section
  \textsuperscript{\emph{c}} See eq.~\ref{eqn:stbasdim}
  \textsuperscript{\emph{d}} Ref.~\citenum{kalea04a}
  \textsuperscript{\emph{e}} Ref.~\citenum{zanca16a}
  \textsuperscript{\emph{f}} Ref.~\citenum{harrr06a}
\end{table}

\section{Economization methods with results}

The most direct method of increasing the fidelity of diatomic binding
in the interacting pairs would be to increase the orbital and
configuration bases going into the diatomic calculations of
eq.~\ref{eqn:diytsolv}, reducing the magnitude of the variational
discrepancy for diatoms at finite and infinite separation and
presumably reducing the difference between those discrepancies.  The
increase in the atomic state basis which would result would be
disadvantageous because of the scaling in the size of the polyatomic
problem (eq.~\ref{eqn:stbasdim}).  At least two alternate approximate
approaches seem worth pursuing: sifting the configuration space for
those expected to be most important to bonding and, at fixed atomic
spectral size, directly injecting improved interaction energies into
the pair matrices.

\subsection{Reducing the size of the atomic spectrum}

It is desirable to prune the atomic spectra (tabs. ~\ref{tbl:hspect}
and~\ref{tbl:cspect}) by selecting or constructing only those states
especially important for molecular binding.  By way of example,
consider the single-excitation, atomic-spectral basis for carbon
labeled ``VB3'' in tab.~\ref{tbl:cspect}) which has a multiplicity of
6230.  The present focus is on describing low-lying polyatomic states
and this search for a pruned atomic basis began with the
presumption that low-lying atomic states are the most important.
Initial efforts therefore eliminated altogether atomic states in the
phase-consistent basis of eq.~\ref{eqn:atpc} with an energy above a
fixed threshold (for example those above the $\mathrm{^3P_g(1)}$,
$\mathrm{^1D_g(1)}$, $\mathrm{^1S_g(1)}$, and $\mathrm{^5S_u}$ states
in carbon).  This was surprisingly ineffective, presumably because of
the total neglect of the above-threshold, valence-excited states.

The next attempt involved performing an initial atomic calculation in
a large configuration set and examining the results of
eq.~\ref{eqn:atsol}.  The Young tableaux supporting complete manifolds
(right-hand side) were sifted based on their population-based
contribution to a set of preselected lowest-energy atomic states
(left-hand-side), for example the four lowest carbon state manifolds.
Those with a contribution exceeding a threshold value were retaining
for a second phase of atomic CI in a reduced configuration set.  By
adjusting the threshold, Young-tableau bases of varying size could be
defined and carried forward into the construction of diatomic
matrices.  This was only partially successful as it seemed to still
relatively devalue the valence-excited states which were not able to
be isolated, but whose spectral density was spread among several of the
higher energy states.

Finally a two-step procedure, completely within the atomic
calculations, was employed to select the atomic-tableau functions.
The orbitals of a large basis were divided into an inner set and an
outer set.  (In the frozen-core carbon example it was natural to put
the 2s and 2p orbitals in the inner set.)  A list of reference valence
tableaux, indexed by r was then assembled from every allowable
occupation of the inner orbitals.  The vector-matrix form of
eq.~\ref{eqn:atsol} is separated by components and every state, s on
the left-hand-side was then assigned a ``valence score,''
$\mathrm{v_s}$, based upon the summed contributions of the reference
configurations to it.

\begin{equation}                                                                
  \mathrm{v_s} = \sum_\mathrm{r} \Phi_\mathrm{r} \mathrm{V_{rs}}     
  \label{eqn:CIOPa}                                                            
\end{equation}

\noindent In a second step each of the tableau in the larger
configuration set, indexed by k, was given a composite score,
$\mathrm{t_k}$: the product of the contribution of the tableau to a
state and the valence score of that state, summed over all the atomic
states.

\begin{equation}                                                                
  \mathrm{t_k} = \sum_\mathrm{s} \mathrm{v_s} \mathrm{V_{ks}}     
  \label{eqn:CIOPb}                                                            
\end{equation}

\noindent This would quantify the relative importance of individual
tableau, but as the Spectral Theory is founded upon using whole atomic
manifolds (such as those listed, {\it e.g} in eqs.~\ref{eqn:cbasex}
and~\ref{eqn:hbasex}), a further manipulation is required.  The
tableaux can be partitioned into sets, indexed by q, which together
support distinct manifolds.  The collective importance of each set,
$\mathrm{M_q}$ in number with each indexed by $\mathrm{k_q}$ can then
be quantified by averaging the tableau weight over the set:

\begin{equation}                                                                
  \mathrm{t_q} = {1\over \mathrm{M_q}} \sum_\mathrm{k_q} \mathrm{t_{kq}}     
  \label{eqn:CIOPb}                                                            
\end{equation}

\noindent Sets of tableaux with an averaged score above an adjustable threshold
were then excluded or retained and this procedure provided
configuration sets of varying size and fidelity to the original set.
In this way, the variable valence-content of excited states could be
flexibly accounted for, even while focusing on the lower, more
energy-accessible states presumably most important for binding.  For
nonhydrogen atoms this procedure also allows a fine-tuning in the size
of the atomic-state bases intermediate between the multiplicities
(Tab.~\ref{tbl:cspect}) which arise directly from the orbitals
employed and the degree of excitation.  Note that this basis reduction
is propagated to the polyatomic dimension because of its polynomial
scaling in the number of atomic states (eq.~\ref{eqn:stbasdim}).

Table~\ref{tbl:cspect} shows term energies for two of these bases
formed by selecting tableaux from all 6230 of those of the
single-excitation VB3 calculation.  That labeled ``A'' is obtained
with a set threshold of 0.07, while the larger basis ``B'' uses 0.01.
By way of example, basis ``A'' adds to the 70 states of
eq.~\ref{eqn:cbasex}, 168 states of 22 new manifolds of term
symmetries:

\begin{equation}
\mathrm{^5P_g}, \mathrm{^3P_g}, \mathrm{^3P_u},
\mathrm{^1P_u}, \mathrm{^1P_g}, \mathrm{^3D_g},
\mathrm{^3F_u}, \mathrm{^3S_g}, \mathrm{^1D_g},
\mathrm{^3D_g}, \mathrm{^1D_u},  \\
\mathrm{^3P_u},
\mathrm{^3D_u}, \mathrm{^1F_u}, \mathrm{^3S_g},
\mathrm{^1P_u}, \mathrm{^3P_g}, \mathrm{^3P_g},
\mathrm{^1S_g}, \mathrm{^1P_g}, \mathrm{^1D_g}, \mathrm{^1S_g}
\label{eqn:cbasexAX}
\end{equation}

\noindent By making the threshold value sufficiently large, the
valence-excited basis is recovered; setting it very low will give back
the original, full single-excitation basis.

\begin{figure}
  \includegraphics[width=1.0\textwidth]{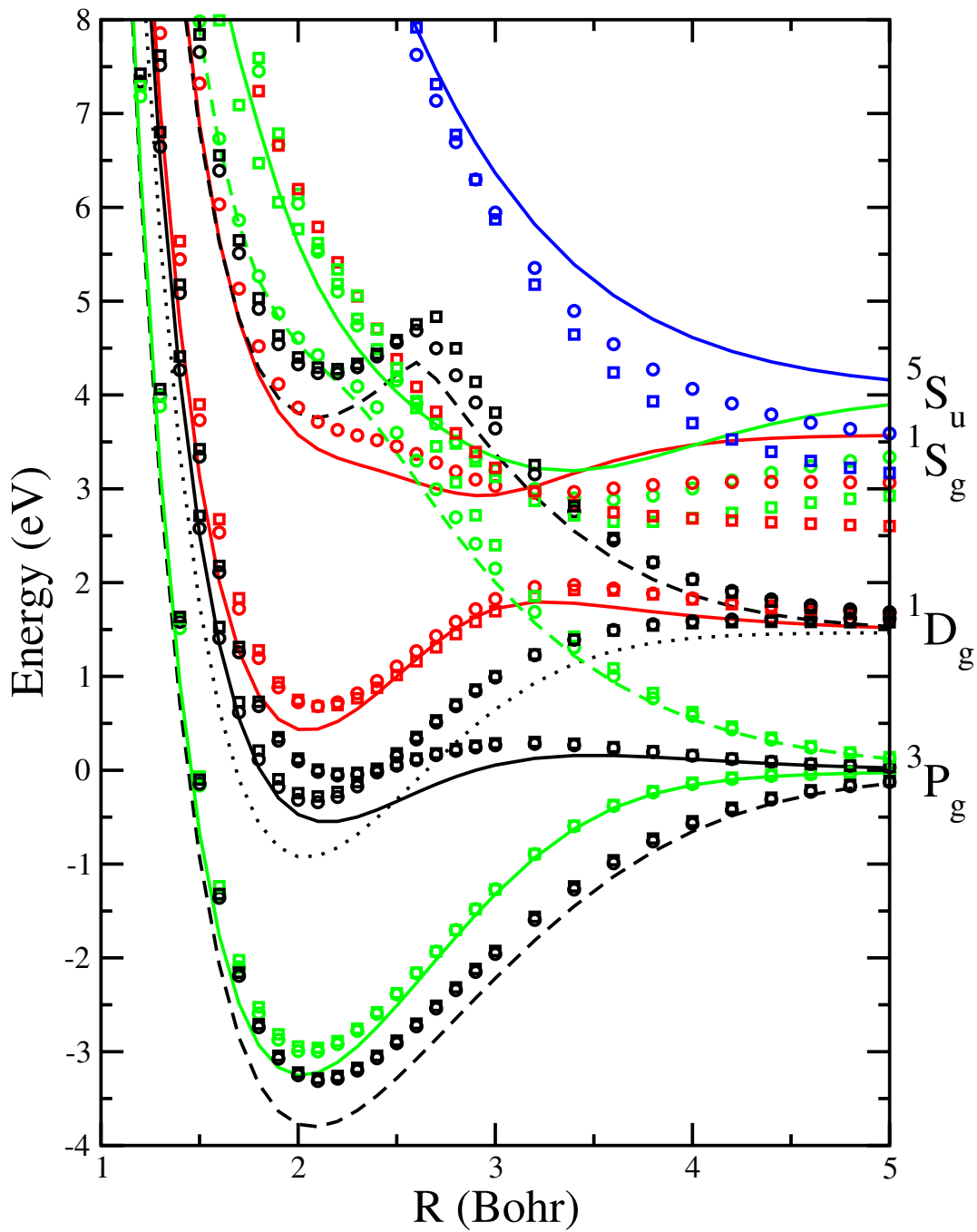}
  \caption{CH interaction energies with reduced carbon-atom spectral basis.
    Lines are identical to the thinner lines of Fig.~\ref{fgr:chPES},
    VB3, single excitation (same color and line dash pattern and
    labeled with asymptotic carbon state).
    Circles-basis ``A''; squares-basis ``B''; same color scheme as
    corresponding lines.}
  \label{fgr:chPESb}
\end{figure}

Figure~\ref{fgr:chPESb} displays diatomic CH results for these reduced
bases.  For comparison the lines redisplay the single-excitation
results of the VB3 basis from Fig.~\ref{fgr:chPES} (same color and
line dash pattern so that symmetries can be assigned).  In
corresponding colors, and often difficult to distinguish, circles and
squares show the results for the VB3 hydrogen basis paired with bases
``A'' and ``B.''  This procedure seems only partially successful in
decreasing the size of the atomic spectrum while yielding sufficiently
bound diatomic wells.


\subsection{Injecting improved interaction energies}

This means of improving the description of bonding in molecular
aggregates for fixed atomic-spectral dimension can most easily be
understood with reference to eq.~\ref{eqn:distsol}.  For a relatively
small basis, the diatomic matrix
${\bf H}_{\bf d}^{(\alpha,\beta)}(R_{\alpha\beta})$ for each of the
archived values of $R_{\alpha\beta}$ can be diagonalized, yielding
${\bf V}_{{\bf H}_{\bf d}}^{(\alpha,\beta)}(R_{\alpha\beta})$ and the
interaction-energy eigenvalues along the diagonal of
${\bf E}^{(\alpha,\beta)}(R_{\alpha\beta})$.  Selected, in this case
low-energy, eigenvalues can then be replaced with values calculated
using a larger configuration set (and shifted to reflect the
difference in atomic ground-state energies).  This combines a superior
treatment of binding in the low-energy states with a reduced, but
still complete (in the smaller basis), description of the excited
states.  The inverse of the original
${\bf V}_{{\bf H}_{\bf d}}^{(\alpha,\beta)}(R_{\alpha\beta})$ is
right-multiplied on each side of eq.~\ref{eqn:distsol} yielding new
``hybrid basis'' diatomic matrices.

\begin{figure}
  \includegraphics[width=1.0\textwidth]{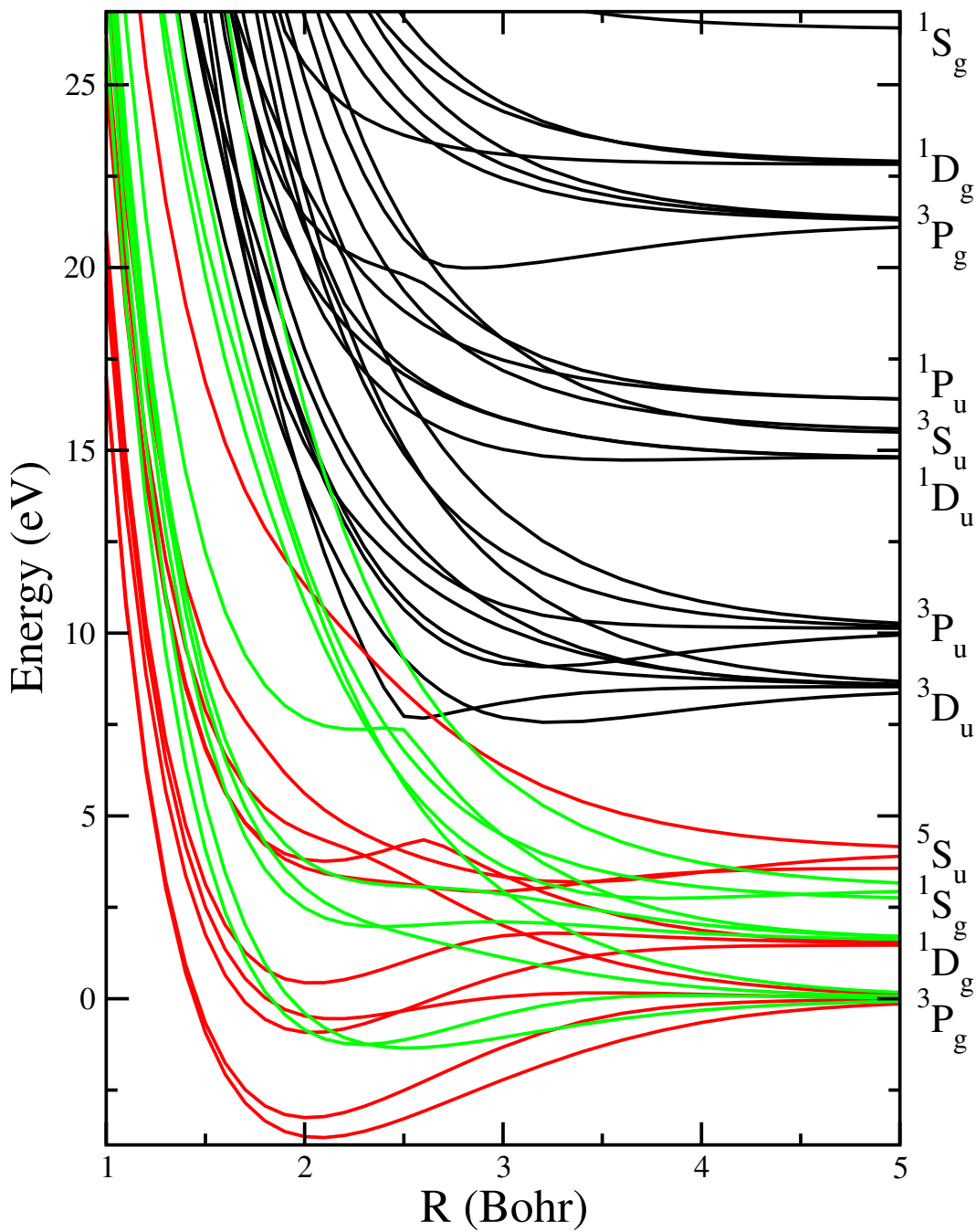}
  \caption{CH potential energy curves relevant for construction of the
    ``A'' fused-diatomic basis.  For states dissociating to the lowest
    four asymptotes, the red curves (single-excitation VB3) are
    substituted for their valence-excitation analogues, the green
    curves to provide a basis with the same size as the no-excitation
    VB3, but increased binding in the low-energy states.}
  \label{fgr:chPESc}
\end{figure}

Figure~\ref{fgr:chPESc} illustrates in more detail the components of
these fused archive matrices.  The black and green curves are all the
diatomic curves which result from the smaller calculation with only
valence excitation within the VB3 basis.  The green curves are the
(relatively underbound) curves already displayed in
Fig.~\ref{fgr:chPES} as the thicker lines.  The black curves are all
the remaining diatomic states in this basis and have not been
displayed heretofore.  The significant break between the asymptotic
energies of these two sets of curves should be noted and compared to
the carbon-atom state energies of tab.~\ref{tbl:cspect}.

The red curves are the diatomic results for states which dissociate to
the four lowest carbon-atom asymptotes for single-excitation out of
the valence reference space using the same basis.  These were
previously displayed in Fig.~\ref{fgr:chPES} as the thinner lines.
Substituting the red curves in preference to their green counterparts
gives increased binding in the lower diatomic states even if it may
distort some of the higher interactions.  (Note especially the avoided crossing
prominent near R=2.5 Bohr.)  In tables~\ref{tbl:X2BE}
and~\ref{tbl:CHnBE} the basis which results from this procedure is
labeled ``C.''  Basis ``D'' is obtained from combining the same lower
diatomic curves with the higher curves of the carbon basis labeled
``A'' in table~\ref{tbl:cspect} and described in the previous section.


\begin{figure}
  \includegraphics[width=1.0\textwidth]{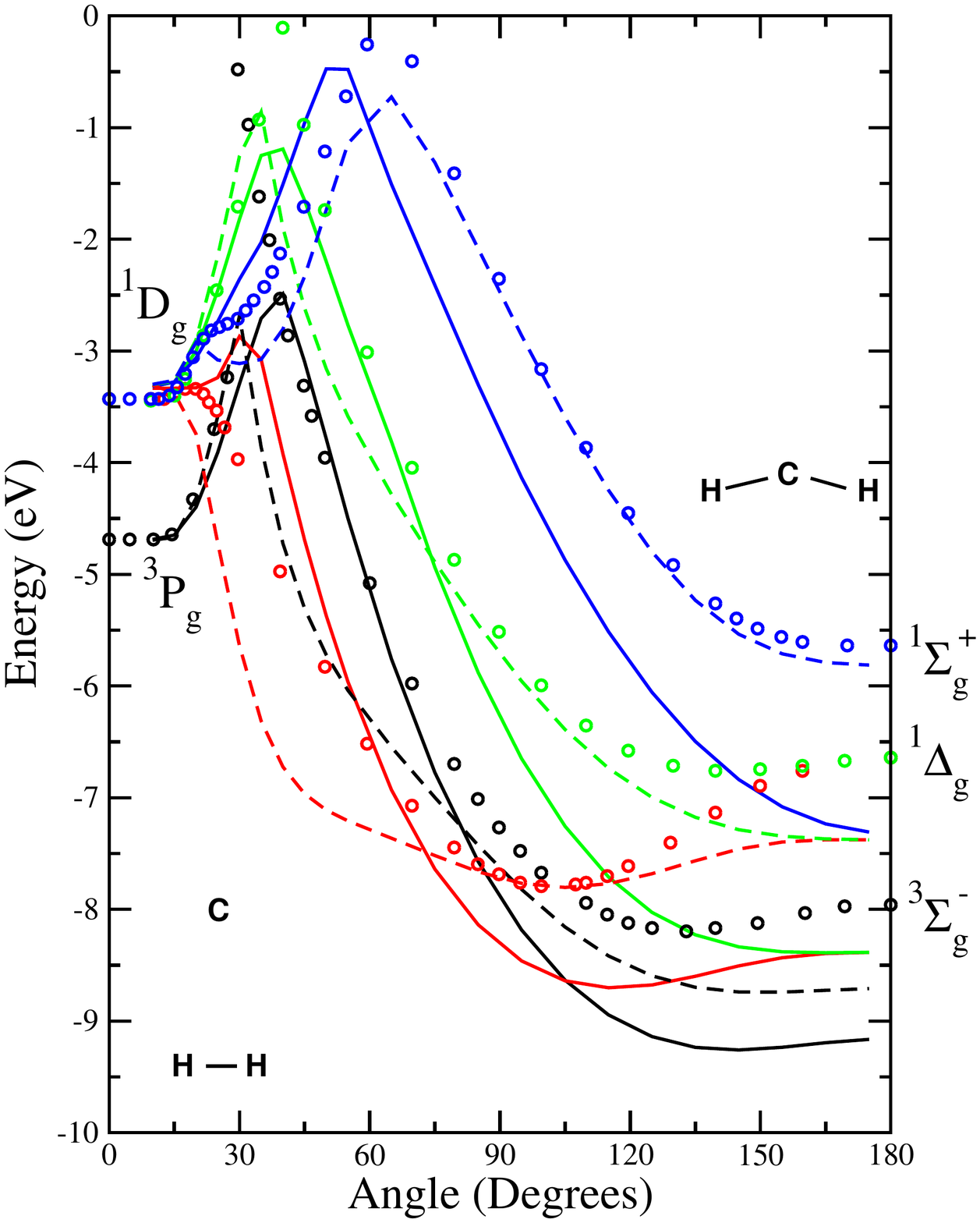}
  \caption{Optimized $\mathrm{CH_2}$($\mathrm{C_{2v}}$) binding
    energies for fixed angles (same optimization procedure, state
    colors and asymptotic channels as in Fig.~\ref{fgr:ch2PES}).
    Solid lines-Spectral Theory using basis ``C''; dashed
    lines-Spectral Theory using basis ``D''; circles-large-basis MRCI
    results, ref.~\citenum{kalea04a}.}
  \label{fgr:ch2PESb}
\end{figure}

Figure~\ref{fgr:ch2PESb} shows results for $\mathrm{CH_2}$, repeating
the optimization procedure of Fig.~\ref{fgr:ch2PES} for bases ``C''
(solid lines) and ``D'' (dashed lines) (along with the same reference,
high-level theoretical results).  The injection of diatomic curves
from higher-level calculations might seem somewhat artificial
(especially if pursued too aggressively) but at least in this example
does not seem to result in catastrophic distortion of the polyatomic
results.  These and the results in table~\ref{tbl:CHnBE} illustrate
that fairly economical bases can furnish results in substantial
agreement with more expensive theoretical methods.


\section{Continuing developments and conclusions}

The finite-basis, pairwise-antisymmetrized form of the Spectral Theory
of chemical bonding has been described and applied to the ground and
low-excited states of some prototypical small hydrocarbons to
illustrate its characteristics.  Progress has also been reported on
approximations to minimize the number of atomic states necessary to
faithfully describe atomic and diatomic contributions.  As the total
polyatomic basis dimension is polynomial in atomic-state number but
exponential in the number of atoms, application of this theory to
larger molecules will require further attention to breaking the
exponential scaling.  In particular, future reports will describe
economizing approximations inspired by density-matrix renormalization
group theory\cite{whits92a,whits92b,whits93a} and utilizing some
unique simplifications which result in application of these methods to
pairwise Hamiltonians (eq.~\ref{eqn:polydimat}).

\section{Acknowledgments}

I would like to acknowledge Peter W. Langhoff for valuable discussions
concerning this work and related approaches, Jerry Boatz for his ready
ear and helpful comments, Wayne Kalliomaa and Stefan Schneider for
their supporting encouragement, and Mike Berman for significant early
funding.

\bibliography{JPC2022}

\end{document}